\documentclass[12pt]{article}
\usepackage{epsfig}
\usepackage{amssymb}
\usepackage{amsmath}
\usepackage{amsfonts}
\usepackage{graphicx}
\usepackage{mathrsfs}
\usepackage[dvips]{color}
\usepackage{multirow}

\newtheorem{theorem}{Theorem}

\newcommand{\bsigma}{\boldsymbol{\sigma}}

\newcommand{\R}{\mathbb{R}}
\newcommand{\C}{\mathbb{C}}

\newcommand{\fn}{{\mathfrak{n}}}
\newcommand{\ft}{\mbox{\large$\mathfrak{t}$}}
\newcommand{\fr}{\mbox{\large$\mathfrak{r}$}}
\newcommand{\fs}{\mathfrak{s}}

\newcommand{\fz}{\mathfrak{z}}

\newcommand{\fD}{\mathfrak{D}}

\newcommand{\fK}{\mathfrak{K}}

\newcommand{\fS}{\mathfrak{S}}

\newcommand{\bba}{\mathbf{a}}
\newcommand{\bb}{\mathbf{b}}

\newcommand{\bA}{\mathbf{A}}

\newcommand{\bB}{\mathbf{B}}

\newcommand{\bI}{\mathbf{I}}

\newcommand{\bM}{\mathbf{M}}
\newcommand{\bN}{\mathbf{N}}

\newcommand{\bS}{\mathbf{S}}

\newcommand{\bfeta}{\boldsymbol{\eta}}

\newcommand{\cE}{\mathcal{E}}

\newcommand{\cN}{\mathcal{N}}

\newcommand{\cP}{\mathcal{P}}

\newcommand{\cS}{\mathcal{S}}
\newcommand{\cT}{\mathcal{T}}
\newcommand{\cU}{\mathcal{U}}
\newcommand{\bcU}{\boldsymbol{\cU}}

\newcommand{\be}{\begin{equation}}
\newcommand{\ee}{\end{equation}}
\newcommand{\bea}{\begin{eqnarray}}
\newcommand{\eea}{\end{eqnarray}}
\newcommand{\nn}{\nonumber}
\newcommand{\kt}{\rangle}
\newcommand{\br}{\langle}

\newcommand{\ed}{\end{document}}
\newcommand{\pbr}{\prec}
\newcommand{\pkt}{\succ}

\newcommand{\bi}{\begin{itemize}}
\newcommand{\ei}{\end{itemize}}

\newcommand{\bce}{\begin{center}}
\newcommand{\ece}{\end{center}}

\newcommand{\sH}{\mathscr{H}}

\newcommand{\sS}{\mathscr{S}}

\newcommand{\RE}{{\rm Re}}
\newcommand{\IM}{{\rm Im}}

\newcommand{\bigchi}{\mbox{\large$\chi$}}

\begin{document}

%\title{%A Global Approach to
%Scattering Theory and $\cP\cT$-Symmetric Scattering Systems %in One Dimension
%}

\title{%A Global Approach to
Scattering Theory and $\cP\cT$-Symmetry %in One Dimension
}

\author{Ali~Mostafazadeh\thanks{E-mail address:
amostafazadeh@ku.edu.tr}\\[6pt]
Departments of Mathematics and Physics, Ko\c{c} University,\\  34450
Sar{\i}yer, Istanbul, Turkey}

\date{ }
\maketitle

\begin{abstract}

We outline a global approach to scattering theory in one dimension that allows for the description of a large class of scattering systems and their $\mathcal{P}$-, $\mathcal{T}$-, and $\mathcal{P}\mathcal{T}$-symmetries. In particular, we review various relevant concepts such as Jost solutions, transfer and scattering matrices, reciprocity principle, unidirectional reflection and invisibility, and spectral singularities. We discuss in some detail the mathematical conditions that imply or forbid reciprocal transmission, reciprocal reflection, and the presence of spectral singularities and their time-reversal. We also derive generalized unitarity relations for time-reversal-invariant and $\mathcal{P}\mathcal{T}$-symmetric scattering systems, and explore the consequences of breaking them. The results reported here apply to the scattering systems defined by a real or complex local potential as well as those determined by energy-dependent potentials, nonlocal potentials, and general point interactions.
\end{abstract}

{\small
\tableofcontents}

\section{Basic setup for elastic scattering in one dimension}

The theory of the scattering of waves by obstacles or the interactions
modelling them rests on the assumption that the strength of the
interaction diminishes at large distances from the source of the
incident wave and the detectors of the scattering experiment, so
that in the vicinity of the source and detectors the wave can be
safely approximated by a plane wave. A consistent implementation of
this assumption requires the existence of solutions of the relevant
wave equation that tend to plane waves at spatial infinities. For a
time-harmonic scalar wave, $e^{-i\omega t}\psi(x)$, propagating in
one dimension, this requirement takes the form of the following
asymptotic boundary conditions:
    \be
    \psi(x)\to\left\{\begin{array}{ccc}
    A_-(k)e^{ikx}+B_-(k)e^{-ikx} & {\rm for}& x\to-\infty,\\
    A_+(k)e^{ikx}+B_+(k)e^{-ikx} & {\rm for}& x\to+\infty,
    \end{array}\right.
    \label{am-pw}
    \ee
where $A_\pm(k)$ and $B_\pm(k)$ are complex-valued functions of the
wavenumber $k$, which we take to be a positive real variable unless
otherwise is clear. The factors $e^{ikx}$ and $e^{-ikx}$ appearing in
(\ref{am-pw}) are related to the solutions, $e^{i(kx-\omega t)}$ and
$e^{-i(kx+\omega t)}$, of the wave equation in the absence of the interaction.
They represent the right- and left-going waves, respectively.

As a principal example, consider the scattering phenomenon described by the
Schr\"odinger equation,
    \be
    -\psi''(x)+v(x)\psi=k^2\psi(x),
    \label{am-sch-eq}
    \ee
where $v(x)$ is a real or complex interaction potential. The existence of the solutions of this equation that satisfy (\ref{am-pw}) restricts the rate at which $|v(x)|$ decays to zero as $x\to\pm\infty$. We can also consider the more general situations
where the potential is energy-dependent. For example consider the
Helmholtz equation,
    \be
    \psi''(x)+k^2\hat\varepsilon(x,k)\psi(x)=0,
    \label{am-helm-eq}
    \ee
which describes the interaction of polarized electromagnetic waves
having an electric field of the form $E_0 e^{-i\omega t}\psi(x)$ pointing
along the $y$-axis with an isotropic nonmagnetic media represented
by a real or complex relative permittivity profile $\hat\varepsilon(x,k)$, \cite{born-wolf}.
We can express (\ref{am-helm-eq}) in the
form (\ref{am-sch-eq}) provided that we identify $v(x)$ with the
energy-dependent optical potential:
    \be
    v(x,k)=k^2[1-\hat\varepsilon(x,k)].
    \label{am-opt-pot}
    \ee

The scattering setup we have outlined above also applies for the scattering of
waves described by nonlocal and nonlinear Schr\"odinger equations \cite{am-muga,ruschhaupt,am-prl-2013,pla-2017},
    \bea
    &&-\psi''(x)+\int_{-\infty}^\infty V(x,x')\psi(x')dx'=k^2\psi(x),
    \label{am-NLoc-sch-eq} \\
    && -\psi''(x)+V(x,\psi(x))\psi(x)=k^2\psi(x),
    \label{am-NL-sch-eq}
    \eea
if the nonlocal and nonlinear potentials, $V(x,x')$ and $V(x,\psi(x))$ decay sufficiently rapidly as $x\to\pm\infty$ so that (\ref{am-NLoc-sch-eq}) and (\ref{am-NL-sch-eq}) admit solutions satisfying (\ref{am-pw}). This is clearly the case for a confined nonlocal and nonlinear interactions \cite{devillard,am-prl-2013}, where
    \begin{align*}
    &V(x,x')=v(x)\delta(x-x')+F(x,x')\bigchi_{[a,b]}(x),
    &&V(x,\psi(x))=v(x)+F(x,\psi(x))\bigchi_{[a,b]}(x),
    \end{align*}
$\delta(x)$ stands for the Dirac delta function, $F$ is a complex-valued function of a pair of real or complex variables, $[a,b]$ is a closed interval of real numbers,
    \[\bigchi_{[a,b]}(x):=\left\{\begin{array}{ccc}
    1 & {\rm for} & x\in[a,b],\\
    0 & {\rm for} & x\notin[a,b],\end{array}\right.\]
and we use the symbol ``$:=$'' (respectively ``$=:$'') to state that the right-hand (respectively  left-hand) side is the definition of the left-hand (respectively right-hand.)

Another class of scattering problems that we can treat using our general framework for scalar-wave scattering in one dimension is that of single- or multi-center point interactions \cite{am-jpa-2011}. These correspond to scalar waves $\psi(x)$ that satisfy
    \be
    \begin{aligned}
    &-\psi''(x)=k^2\psi(x)&&{\rm for}
    \quad\quad x\in\R\setminus\{c_1,c_2,\cdots,c_n\},\\
    &\left[\begin{array}{c}
    \psi(c_j^+)\\
    \psi'(c_j^+)\end{array}\right]=\bB_j\left[\begin{array}{c}
    \psi(c_j^-)\\
    \psi'(c_j^-)\end{array}\right] \quad\quad&&{\rm for}\quad\quad j\in\{1,2,\cdots,n\},
    \end{aligned}
    \label{am-pt-int}
    \ee
where $c_1,c_2,\cdots,c_n$ are distinct real numbers representing the interaction centers, for every function $\phi(x)$ the symbols $\phi(c_j^-)$ and $\phi(c_j^+)$ respectively denote the left and right limit of $\phi(x)$ as $x\to c_j$, i.e., $\phi(c_j^\pm):=\lim_{x\to c_j^\pm}\phi(x)$, and $\bB_j$ are possibly $k$-dependent $2\times 2$ invertible matrices. The point interactions of this type may be used to model electromagnetic interface conditions \cite{am-ap-2016a}.

The best known example of a single-center point interaction is the delta-function potential $v(x)=\fz\,\delta(x)$ with a coupling constant $\fz$. It corresponds to the choice: $n=1$, $c_1=0$, and
    \be
    \bB_1=\left[\begin{array}{cc}
    1 & 0\\
    \fz & 1\end{array}\right].
    \label{am-pt-int-delta}
    \ee

In a scattering experiment, the incident wave is emitted by its
source which is located at one of the spatial infinities
$\pm\infty$, and the scattered wave is received by the detectors
which are placed at one or both of these infinities. If the source
is located at $-\infty$, the incident wave travels towards the
region of the space where the interaction has a sizable strength. A
part of it passes through this region and reaches the detector at
$+\infty$. The other part gets reflected and travels towards the
detector at $-\infty$. As a result, the incident and transmitted
waves are right-going while the reflected wave is left-going. This
scenario is described by a solution $\psi_l(x)$ of the wave equation
that has the following asymptotic behavior.
    \be
    \psi_l(x)\to\left\{\begin{array}{ccc}
    \cN\left[e^{ikx}+\fr_{l}(k)\,e^{-ikx}\right] & {\rm for}& x\to-\infty,\\
    \cN\, \ft_l(k)\,e^{ikx} & {\rm for}& x\to+\infty,
    \end{array}\right.
    \label{am-jost-left}
    \ee
where $\cN$ is the amplitude of the incident wave, and $\fr_l(k)$ and
$\ft_l(k)$ are complex-valued functions of $k$ that are respectively
called the {\em left reflection and transmission amplitudes}.
Similarly, we have the solution $\psi_r(x)$ of the wave equation
that corresponds to the scattering of an incident wave that is
emitted from a source located at $x=+\infty$. This satisfies
    \be
    \psi_r(x)\to\left\{\begin{array}{ccc}
    \cN\, \ft_r(k)\,e^{-ikx} & {\rm for}& x\to-\infty,\\
    \cN \left[e^{-ikx} + \fr_r(k)\,e^{ikx}\right] & {\rm for}& x\to+\infty,
    \end{array}\right.
    \label{am-jost-right}
    \ee
where $\fr_r(k)$ and $\ft_r(k)$ are respectively the {\em right
reflection  and transmission amplitudes}.

Scattering experiments involve the measurement of the reflection and
transmission amplitudes, $\fr_{l/r}(k)$ and $\ft_{l/r}(k)$, or their
modulus square, $|\fr_{l/r}(k)|^2$ and $|\ft_{l/r}(k)|^2$, which are
respectively called the {\em left/right reflection and transmission
coefficients}\footnote{These are occasionally labelled by $T^{l/r}(k)$
and $R^{l/r}(k)$, \cite{am-muga,am-prl-2013}. Here we refrain from
using this notation, because some references use these symbols for
the reflection and transmission amplitudes and not their
modulus squared \cite{am-lin-2011}.}. By solving a scattering problem we mean
the determination of $\fr_{l/r}(k)$ and $\ft_{l/r}(k)$. We sometimes call these the ``{\em scattering data}''.

If $\fr_{l/r}(k_0)=0$ for some wavenumber $k_0\in\R^+$, we say that the scatterer\footnote{By a scatterer we mean the interaction causing the propagation of a wave differ from that of a plane wave.} is {\em reflectionless from the left/right} or simply {\em left/right-reflectionless} at $k=k_0$. Similarly we call it {\em left/right transparent} at $k=k_0$, if $\ft_{l/r}(k_0)=1$. A scatterer is invisible from the left or right if it is both reflectionless and transparent from that direction. In this case we call it {\em left/right-invisible}. {\em Unidirectional reflectionlessness} (respectively {\em unidirectional invisibility}) refers to situations where a scatterer is reflectionless (respectively invisible) only from the left or right \cite{am-lin-2011}. The reflectionlessness, transparency, and invisibility of a scatterer are said to be {\em broadband} if they hold for a finite or infinite interval of positive real values of $k$.

If the wave equation is linear, we can scale $\psi_{l/r}$ and  work
with $\psi_{+/-}:=\psi_{l/r}/\cN \ft_{l/r}$. These satisfy:
    \be
    \begin{aligned}
    \psi_{\pm}(x) \to &~ e^{\pm ikx} &&{\rm for}~~x\to\pm\infty,\\
    \psi_{+}(x) \to &~\frac{1}{\ft_l(k)}\,e^{ikx}+\frac{\fr_l(k)}{\ft_l(k)}\,e^{-ikx}
    && {\rm for}~~x\to-\infty,\\
    \psi_{-}(x) \to &~\frac{\fr_r(k)}{\ft_r(k)}\,e^{ikx}+\frac{1}{\ft_r(k)}\,e^{-ikx}
    && {\rm for}~~x\to+\infty,
    \end{aligned}
    \label{am-jost1}
    \ee
and are called the {\em Jost solutions}. It turns  out
that the Schr\"odinger equation~(\ref{am-sch-eq}) admits Jost
solutions, if $\int_{-\infty}^{\infty}\sqrt{1+x^2}\,|v(x)|
dx<\infty$. This is equivalent to the {\em Faddeev condition}:
    \be
    \int_{-\infty}^{\infty}(1+|x|)|v(x)| dx<\infty.
    \label{am-faddeev-condi}
    \ee
Under this condition the Jost solutions exist not only for real and positive values of $k$, but also for complex values of $k$ belonging to the upper-half complex plane, i.e., $k\in\{\fz\in\C~|~\IM(\fz)\geq 0\}$. Furthermore, in this half-plane they are continuous functions of $k$, \cite{am-kemp-1958}.

Faddeev condition clearly holds for {\em finite-range potentials} which vanish
outside a finite interval (have a compact support), and {\em exponentially decaying
potentials} which satisfy
    \be
    e^{\mu_\pm |x|}|v(x)|<\infty~~~{\rm for}~~~x\to\pm\infty,
    \label{am-exp-decay}
    \ee
for some $\mu_\pm\in\R^+$. Notice that finite-range potentials fulfill this condition for all $\mu_\pm\in\R^+$. Therefore they share the properties of exponentially decaying potentials that follow from (\ref{am-exp-decay}).

In this article we use the term ``{\em scattering potential }'' for real or complex-valued potentials $v(x)$ that satisfy the Faddeev condition~(\ref{am-faddeev-condi}).

\section{Transfer matrix}
\label{am-TM}

Consider a linear scalar wave equation that admits time-harmonic solutions $e^{-i\omega t}\psi(x)$ fulfilling the asymptotic boundary conditions~(\ref{am-pw}). We can identify these solutions by the pairs of column vectors:
    \[ \left[\begin{array}{c} A_-(k)\\ B_-(k)\end{array}\right]
    \quad{\rm and}\quad
    \left[\begin{array}{c} A_+(k)\\ B_+(k)\end{array}\right].\]
The  $2\times 2$ matrix $\bM(k)$ that connects these is called the {\em transfer matrix}
\cite{am-razavy, am-sanchez-soto}. By definition, it satisfies
    \be
    \bM(k)\left[\begin{array}{c} A_-(k)\\ B_-(k)\end{array}\right]=
    \left[\begin{array}{c} A_+(k)\\ B_+(k)\end{array}\right].
    \label{am-M-def}
    \ee
If we demand that the knowledge of the solution of the wave equation at either of the spatial infinities, $x\to\pm\infty$,  determines it uniquely, $\bM(k)$ must be invertible. In what follows we assume that this is the case, i.e., $\det\bM(k)\neq 0$.\footnote{In Section~\ref{am-S4}, we prove that this conditions holds for the scattering systems described by the Schr\"odinger eqution (\ref{am-sch-eq}).}

We can express the entries of $\bM(k)$ in terms of the  reflection
and transmission amplitudes by implementing (\ref{am-M-def}) for the
Jost solutions. This requires the identification of the coefficient functions
$A_\pm(k)$ and $B_\pm(k)$ for $\psi(x)=\psi_\pm(x)$. Comparing
(\ref{am-pw}) and (\ref{am-jost1}), we see that for
$\psi(x)=\psi_+(x)$,
    \begin{align}
    &A_-=\frac{1}{\ft_l}, && B_-=\frac{\fr_l}{\ft_l}, && A_+=1, && B_+=0.
    \label{am-psi-plus}
    \end{align}
Here and in what follows we occasionally suppress the $k$-dependence of $A_\pm(k)$, $B_\pm(k)$, $\fr_{l/r}(k)$, $\ft_{l/r}(k)$, $\bM(k)$, and other relevant quantities for brevity.
Similarly for $\psi(x)=\psi_-(x)$, we have
    \begin{align}
    &A_-=0, && B_-=1, && A_+=\frac{\fr_r}{\ft_r}, && B_+=\frac{1}{\ft_r}.
    \label{am-psi-minus}
    \end{align}
Substituting (\ref{am-psi-plus}) and (\ref{am-psi-minus}) in (\ref{am-M-def}) gives
    \begin{align}
    &\frac{1}{\ft_l}\,\bM\left[\begin{array}{c} 1 \\ \fr_l
    \end{array}\right]=\left[\begin{array}{c} 1 \\ 0
    \end{array}\right],
    &&\bM\left[\begin{array}{c} 0 \\ 1
    \end{array}\right]=\frac{1}{\ft_r}
    \left[\begin{array}{c} \fr_r \\ 1
    \end{array}\right].
    \label{am-MM2}
    \end{align}
The second of these equations implies
    \begin{align}
    & M_{12}=\frac{\fr_r}{\ft_r},
    && M_{22}=\frac{1}{\ft_r}.
    \label{am-M12-M22}
    \end{align}
Using these relations in the first equation in (\ref{am-MM2}), we find
    \begin{align}
    &M_{11}=\ft_l-\frac{\fr_l\fr_r}{\ft_r},
    &&M_{21}=-\frac{\fr_l}{\ft_r}.
    \label{am-M11-M21}
    \end{align}
In view of (\ref{am-M12-M22}) and (\ref{am-M11-M21}),
    \be
    \bM=\frac{1}{\ft_r}\,\left[\begin{array}{cc}
    \ft_l\ft_r-\fr_l\fr_r & \fr_r\\
    -\fr_l & 1\end{array}\right].
    \label{am-M=001}
    \ee
In particular,
    \be
    \det\bM= \frac{\ft_l}{\ft_r}.
    \label{am-det-M=}
    \ee

We can also solve (\ref{am-M12-M22}) and (\ref{am-M11-M21}) for the reflection and
transmission amplitudes in terms of $M_{ij}$. The result is
    \begin{align}
    &\fr_l=-\frac{M_{21}}{M_{22}}, &&\ft_l=\frac{\det\bM}{M_{22}},
    &&\fr_r=\frac{M_{12}}{M_{22}}, &&\ft_r=\frac{1}{M_{22}}.
    \label{am-rt=M}
    \end{align}
Equations~(\ref{am-M=001}) and (\ref{am-rt=M}) show that {\em the knowledge of the transfer matrix is equivalent to solving the scattering problem}. It is also instructive to make the $k$-dependence of the Jost solutions explicit and note that in light of (\ref{am-rt=M}) and (\ref{am-jost1}) their asymptotic expression takes the form
     \be
        \begin{aligned}
        \psi_{\pm}(k,x) \to &~ e^{\pm ikx} &&{\rm for}~~x\to\pm\infty,\\
        \psi_{+}(k,x) \to &~\det\bM(k)^{-1}\left[M_{22}(k)\,e^{ikx}-M_{21}(k)\,e^{-ikx}\right]
        && {\rm for}~~x\to-\infty,\\
        \psi_{-}(k,x) \to &~M_{12}(k)\,e^{ikx}+M_{22}(k)\,e^{-ikx}
        && {\rm for}~~x\to+\infty.
        \end{aligned}
        \label{am-jost2}
        \ee
These relations together with the assumption that $\det\bM(k)\neq 0$
show that {\em as functions of $k$ the entries of the transfer
matrix, $M_{ij}(k)$, have the same analytic properties as the Jost
solutions $\psi_{\pm}(k,x)$}.

A simple consequence of (\ref{am-det-M=}) is that $\det\bM$ is a measure of the violation of reciprocity in transmission; {\em a scattering system has reciprocal transmission if and only if $\det\bM(k)=1$ for all $k\in\R^+$}.

An example of a scattering system that has nonreciprocal transmission is a single-center point interaction~(\ref{am-pt-int}) that is defined by a matching matrix $\bB_1$ with $\det\bB_1\neq 1$, \cite{am-jpa-2011}. To see this, we set $n=1$ and drop the subscript $1$ in $c_1$ and $\bB_1$ in (\ref{am-pt-int}). Clearly for $x\neq c$, every solution of (\ref{am-pt-int}) has the form
    \be
    \psi(x)=A_\pm(k)e^{ikx}+B_\pm(k)e^{-ikx}~~~{\rm for}~~~\pm (x-c)>0.
    \label{am-single-pt-int}
    \ee
We can use this expression to show that
    \be
    \left[\begin{array}{c}
    \psi(c^\pm)\\
    \psi'(c^\pm)\end{array}\right]=\bN_c
    \left[\begin{array}{c}
    A_\pm\\
    B_\pm\end{array}\right],
    \label{am-pt-int-N=}
    \ee
where
    \be
    \bN_c(k):=\left[\begin{array}{cc}
    e^{ick}& e^{-ick}\\
    ik e^{ick} & -ik e^{-ick}\end{array}\right].
    \label{am-pt-int-N=}
    \ee
If we substitute (\ref{am-pt-int-N=}) in (\ref{am-pt-int}), we can relate $A_+(k)$ and $B_+(k)$ to $A_-(k)$ and $B_-(k)$. This gives (\ref{am-M-def}) with the following formula for the transfer matrix of the system.
    \be
    \bM=\bN_c^{-1}\bB\,\bN_c.
    \label{am-pt-int-M=}
    \ee
In particular $\det\bM=\det\bB$.  Therefore, single-center point interactions that satisfy $\det\bB\neq 1$ violate reciprocity in transmission. These are called {\em anomalous point interactions} in \cite{am-jpa-2011}, because they cannot be viewed as singular limits of sequences of scattering potentials.

Next, consider a situation that the solutions $\psi(x)$ of our
linear wave equation have also the form of a plane wave in a closed
interval, $[x_1,x_1+\epsilon]$, where $x_1\in\R$ and
$\epsilon\in\R^+$, i.e., there are coefficient functions $A_1(k)$
and $B_1(k)$ such that for all $x\in[x_1,x_1+\epsilon]$,
    \be
    \psi(x)=A_1(k)e^{ikx}+B_1(k)e^{-ikx}.
    \label{am-mid}
    \ee
In the limit $\epsilon\to 0$ this is certainly true for any $x_1$,
because we can satisfy (\ref{am-mid}) for $x\to x_1$ by setting
    \begin{align}
    &A_1(k)=\frac{e^{-ikx}}{2}\left[\psi(x_1)+\frac{\psi'(x_1)}{ik}\right],
    &&B_1(k)=\frac{e^{ikx}}{2}\left[\psi(x_1)-\frac{\psi'(x_1)}{ik}\right].
    \label{am-mid-bc}
    \end{align}

We can use $x_1$ to disect the original scattering problem into two
pieces.  First, we consider the case where $\psi(x)$ solves the given
wave equation for all $x<x_1$ and takes the form (\ref{am-mid}) for
$x\geq x_1$. Then the choice~(\ref{am-mid-bc}) for $A_1(k)$ and
$B_1(k)$ ensures the continuity and differentiability of the
resulting wave function, namely
    \be
    \psi_1(x):=\left\{\begin{array}{ccc}
    \psi(x) & {\rm for} & x\leq x_1,\\
    A_1(k)e^{ikx}+B_1(k)e^{-ikx} & {\rm for} & x> x_1,\end{array}\right.
    \label{am-phi-minus}
    \ee
at $x=x_1$. We can therefore view $\psi_1(x)$ as the general solution
of the wave equation with the interaction terms missing for $x>x_1$.
Similarly, we introduce
    \be
    \psi_2(x):=\left\{\begin{array}{ccc}
    A_1(k)e^{ikx}+B_1(k)e^{-ikx} & {\rm for} & x< x_1,\\
    \psi(x) & {\rm for} & x\geq x_1,\end{array}\right.
    \label{am-phi-plus}
    \ee
and identify it with the general  solution of the wave equation with the
interaction terms missing for $x<x_1$. According to (\ref{am-pw}),
(\ref{am-phi-minus}), and (\ref{am-phi-plus}),
    \bea
    \psi_1(x)&\to&\left\{\begin{array}{ccc}
    A_-(k)e^{ikx}+B_-(k)e^{-ikx} & {\rm for}& x\to-\infty,\\
    A_1(k)e^{ikx}+B_1(k)e^{-ikx} & {\rm for}& x\to+\infty,
    \end{array}\right.
    \label{am-pw-phi-minus}\\[6pt]
    \psi_2(x)&\to&\left\{\begin{array}{ccc}
    A_1(k)e^{ikx}+B_1(k)e^{-ikx} & {\rm for}& x\to-\infty,\\
    A_+(k)e^{ikx}+B_+(k)e^{-ikx} & {\rm for}& x\to+\infty.
    \end{array}\right.
    \label{am-pw-phi-plus}
    \eea
We can use these relations  together with the definition of the
transfer matrix to introduce the transfer matrices $\bM_j$ for
$\psi_j(x)$. These fulfil
    \begin{align}
    &\bM_1 \left[\begin{array}{c} A_- \\ B_- \end{array}\right]=
    \left[\begin{array}{c} A_1 \\ B_1 \end{array}\right],
    &&\bM_2\left[\begin{array}{c} A_1 \\ B_1 \end{array}\right]=
    \left[\begin{array}{c} A_+ \\ B_+ \end{array}\right].
    \label{am-M-plus-minus}
    \end{align}
Comparing these equations with (\ref{am-M-def}), we see that the
transfer matrix of the original wave equation is given by
    \be
    \bM =\bM_2 \bM_1 .
    \label{am-compose-2}
    \ee

Now, consider dividing the set of real numbers into $n+1$ intervals:
    \begin{align*}
    &I_1:=(-\infty,a_1], && I_2:=[a_1,a_2], && I_3:=[a_2,a_3], &&
    \cdots, && I_{n}:=[a_{n-1},a_n], && I_{n+1}:=[a_n,\infty),
    \end{align*}
and let $\bM_j$ be the transfer matrix for the scattering of a scalar wave
with interactions confined to $I_j$. Then a repeated use of the argument
leading to (\ref{am-compose-2}) shows that the transfer matrix for the
original scattering problem is given by
    \be
    \bM=\bM_{n+1}\bM_n\bM_{n-1}\cdots\bM_1.
    \label{am-compose}
    \ee
This property, which is known as the {\em composition rule for the
transfer matrices}, allows for reducing the scattering problem with
interactions taking place in an arbitrary region of space to simpler scattering
problems where the interaction is confined to certain intervals.

For example, if the interaction has a finite range, i.e., it seizes to exist
outside an interval $[a,b]$, we can set
    \begin{align*}
    a_j:=a+\frac{(j-1)(b-a)}{n}\quad {\rm for}\quad j=1,2,\cdots,n.
    \end{align*}
In this way, by taking large values for $n$ we can reduce the initial
scattering problem to those whose solution requires solving the wave
equation in small intervals. If the interaction is a smooth function of space,
we can approximate it by a constant in each of these intervals. This in turn
simplifies the calculation of $\bM_j$. We can use the result together with
(\ref{am-compose}) to find an approximate expression for $\bM$. Aside
from the technical problems of multiplying a large number of $2\times 2$
matrices, this provides a simple approach for the solution of
the scattering problem for finite-range linear interactions.

We can easily implement this procedure to solve the scattering problem for a muli-center point interaction~(\ref{am-pt-int}). To do this we label the centers of the point interaction so that $c_1<c_2<\cdots<c_n$ and compute the transfer matrix for single-center point interactions associated with $c_j$. As we explained above this has the form
    \be
    \bM_j=\bN_j^{-1}\bB_j\bN_j,
    \label{am-single-pt-int-j-M=}
    \ee
where $\bN_j$ is given by the right-hand side of (\ref{am-pt-int-N=}) with $c$ changed to $c_j$. We can then determine the transfer matrix of the multi-center point interaction by invoking the composition rule (\ref{am-compose}). The result is
    \be
    \bM=\bN_n^{-1}\bB_n\bN_n\bN_{n-1}^{-1}\bB_{n-1}\bN_{n-1}\cdots
    \bN_1^{-1}\bB_1\bN_1.
    \label{am-multi-pt-int-M=}
    \ee
In particular, we find that
    \be
    \det\bM=\det\bB_1\det\bB_2\cdots\det\bB_n.
    \label{am-det-M=det-Bs}
    \ee
Combing this equation with (\ref{am-det-M=}), we infer that {\em a multi-center point interaction violates reciprocity in transmission if and only if it consists of an odd number of anomalous single-center point interactions}.

Next, consider a multi-delta-function potential
    \be
    v(x)=\epsilon\sum_{j=1}^n\fz_j\delta(x-c_j),
    \label{am-multi-delta}
    \ee
where $\epsilon$ is a nonzero real parameter and $\fz_j$ are possibly complex coupling constants. We can identify this with the multi-center point interaction with matching matrices
    \be
    \bB_j=\left[\begin{array}{cc}
    1 & 0\\
    \epsilon\,\fz_j & 1\end{array}\right].
    \label{am-pt-int-delta-j}
    \ee
Substituting this relation in (\ref{am-multi-pt-int-M=}) we find the transfer matrix $\bM$ for (\ref{am-multi-delta}). This has a unit determinant, because $\det\bB_j=1$ and $\bM$ satisfies (\ref{am-det-M=det-Bs}).

It is not difficult to see that the transfer matrix $\bM$ of the multi-delta-function potential (\ref{am-multi-delta}) and hence its entries are polynomials of degree at most $n$ in the parameter $\epsilon$. In view of (\ref{am-jost2}),  and the fact that $\det\bM=1$, this implies that the same is true of the Jost solutions of the Schr\"odinger equation (\ref{am-sch-eq}) for this potential. This observation shows that if we treat $\epsilon$ as a perturbation parameter and perform an $n$-th order perturbative calculation of the Jost solutions, we obtain their exact expression. In view of (\ref{am-jost1}), this allows for determining the reflection and transmission amplitudes of (\ref{am-multi-delta}). We therefore have the following result.
    \begin{theorem}
    The $n$-th order perturbation theory gives the exact solution of the scattering problem for     multi-delta-function potentials with $n$ centers.
    \label{am-thm-pert}
    \end{theorem}
In fact, a direct analysis shows that $n$-th order perturbation theory gives the exact solution of the Schr\"odinger equation (\ref{am-sch-eq}) for multi-delta-function potentials (\ref{am-multi-delta}),~\cite{am-pra-2012}.

\section{Scattering matrix}
\label{am-SM}

By definition, the scattering operator, which is also known as the scattering matrix, maps the waves traveling toward the interaction region (incoming waves) to those traveling away from it (outgoing waves). In one dimension, the boundary conditions (\ref{am-pw}) at spatial infinities show that the incoming waves have the asymptotic form $A_-(k)e^{ikx}$ (respectively $B_-(k)e^{-ikx}$),  if their source is located at $x=-\infty$ (respectively $x=+\infty$), and the outgoing waves tend to $B_+e^{-ikx}$ as $x\to-\infty$ and $A_+(k)e^{ikx}$ as $x\to+\infty$. In light of these observations, we can quantify the scattering operator by a $2\times 2$ matrix $\bS(k)$ that connects $A_-(k)$ and $B_+(k)$ to $A_+(k)$ and $B_-(k)$. Clearly there are four different ways of doing so, namely
    \be
    \begin{aligned}
    &\bS_1\left[\begin{array}{c}
    A_-\\ B_+\end{array}\right]=
    \left[\begin{array}{c}
    A_+\\ B_-\end{array}\right],
    &&\quad \quad \bS_2\left[\begin{array}{c}
    A_-\\ B_+\end{array}\right]=
    \left[\begin{array}{c}
    B_-\\ A_+\end{array}\right],\\
    & \bS_3\left[\begin{array}{c}
    B_+\\ A_-\end{array}\right]=
    \left[\begin{array}{c}
    A_+\\ B_-\end{array}\right],
    &&\quad \quad \bS_4\left[\begin{array}{c}
    B_+\\ A_-\end{array}\right]=
    \left[\begin{array}{c}
    B_-\\ A_+\end{array}\right].
    \end{aligned}
    \label{am-s-matrices}
    \ee
These correspond to various conventions for defining the $\bS$-matrix in one dimension. It is easy to see that
    \begin{align}
    &\bS_2=\bsigma_1\bS_1, &&\bS_3=\bS_1\bsigma_1, &&\bS_4=\bsigma_1\bS_1\bsigma_1,
    \label{am-Ss-connected}
    \end{align}
where $\bsigma_1$ is the first Pauli matrix,
    \[ \bsigma_1:=\left[\begin{array}{cc} 0 & 1\\ 1 & 0\end{array}\right].\]

Next, let us express the entries of $\bS_1$ in terms of the reflection and transmission amplitudes. To do this, we implement the first equation in (\ref{am-s-matrices}) for the Jost solutions $\psi_\pm(x)$. For $\psi(x)=\psi_+(x)$, $A_\pm$ and $B_\pm$ are given by (\ref{am-psi-plus}). Substituting these in the first equation in (\ref{am-s-matrices})  gives
    \be
    \bS_1\left[\begin{array}{cc}
    1 \\ 0 \end{array}\right]=
    \left[\begin{array}{cc} \ft_l \\ \fr_l \end{array}\right].
    \label{am-S1-2}
    \ee
Similarly for $\psi(x)=\psi_-(x)$, we use (\ref{am-psi-plus}) to obtain
    \be
    \bS_1\left[\begin{array}{cc} 0 \\ 1\end{array}\right]=
    \left[\begin{array}{cc} \fr_r \\ \ft_r \end{array}\right].
    \label{am-S1-1}
    \ee
In view of Equations (\ref{am-S1-2}) and (\ref{am-S1-1}),
    \be
    \bS_1=\left[\begin{array}{cc} \ft_l & \fr_r\\
    \fr_l & \ft_r\end{array}\right].
    \label{am-S1=}
    \ee
This relation together with (\ref{am-Ss-connected}) imply
    \begin{align}
    &\bS_2=\left[\begin{array}{cc} \fr_l & \ft_r\\
    \ft_l  & \fr_r \end{array}\right],
    &&\bS_3=\left[\begin{array}{cc} \fr_r & \ft_l\\
    \ft_r & \fr_l \end{array}\right],
    &&\bS_4=\left[\begin{array}{cc} \ft_r & \fr_l\\
    \fr_r & \ft_l\end{array}\right].
    \label{am-S234=}
    \end{align}
According to Equations~(\ref{am-S1=}) and (\ref{am-S234=}), we can use any of $\bS_1, \bS_2, \bS_3$, and $\bS_4$ to encode the information about the scattering properties of the system. They are therefore physically equivalent. We adopt the convention of identifying the $\bS$-matrix with $\bS_1$, i.e., set
    \be
    \bS:=\left[\begin{array}{cc} \ft_l & \fr_r\\
    \fr_l & \ft_r
    \end{array}\right].
    \label{am-S-matrix=}
    \ee
This choise has the appealing property of reducing to the $2\times 2$ identity matrix $\bI$ in the absence of interactions.

Eigenvalues of the scattering matrix turn out to contain some useful information about the scattering properties of the system. In view of (\ref{am-S-matrix=}), they have the form:
    \be
    \fs_\pm=\frac{\ft_l+\ft_r}{2} \pm\sqrt{\left(\frac{\ft_l-\ft_r}{2}\right)^2+\fr_l\fr_r}.
    \label{am-S-eigenvalues-gen}
    \ee
In particular, whenever $\ft_l=\ft_r=:\ft$,
    \be
    \fs_\pm=\ft \pm\sqrt{\fr_l\fr_r}.
    \label{am-S-eigenvalues}
    \ee

Both the transfer and the $\bS$-matrix contain complete information about the scattering data, but in contrast to the transfer matrix the $\bS$-matrix does not obey a useful composition rule. An advantage of the $\bS$-matrix is
the simplicity of its higher-dimensional, relativistic, and field theoretical generalizations \cite{am-weinberg}.\footnote{A genuine multidimensional generalization of the transfer matrix has been recently proposed
in \cite{am-pra-2016}.}

\section{Potential scattering, reciprocity theorem, and invisibility}
\label{am-S4}

Consider the time-independent Schr\"odinger equation~(\ref{am-sch-eq}) for a scattering potential $v(x)$. Then this equation admits Jost solutions $\psi_\pm$ and defines a valid scattering problem.   Being solutions of a second order linear homogeneous differential equation, $\psi_\pm$ are linearly independent if and only if their Wronskian, $W(x):=\psi_-(x)\psi_+'(x)-\psi_+(x)\psi_-'(x)$, does not vanish at some $x\in\R$, \cite{am-boyce-diPrima}. In fact, because the Schr\"odinger equation~(\ref{am-sch-eq}) does not involve the first derivative of $\psi$, $W(x)$ is a constant.\footnote{This can be easily checked by differentiating $W(x)$ and using (\ref{am-sch-eq}) to show that $W'(x)=0$.} We can determine this constant using the asymptotic expression (\ref{am-jost1}) for the $\psi_\pm(x)$. Doing this for $x\to-\infty$ and $x\to+\infty$, we respectively find $W(x)=2ik/\ft_l(k)$ and $W(x)=2ik/\ft_r(k)$. This proves the following reciprocity theorem.
    \begin{theorem}[Reciprocity in Transmission]
    \label{am-thm1}
    The left and right transmission amplitudes of every real or complex scattering potential coincide, i.e.,
        \be
    \ft_l(k)=\ft_r(k).
    \label{am-t=t}
    \ee
    \end{theorem}
    
In the following we use $\ft(k)$ for the common value of $\ft_l(k)$ and $\ft_r(k)$ whenever a scattering system has reciprocal transmission.

In view of Equations~(\ref{am-M=001}), (\ref{am-det-M=}), (\ref{am-rt=M}), (\ref{am-S-matrix=}), and (\ref{am-t=t}), the transfer and scattering matrices and the scattering data associated with real or complex  scattering potentials satisfy:
    \begin{align}
    &\bM=\frac{1}{\ft}\,\left[\begin{array}{cc}
    \ft^2-\fr_l\fr_r & \fr_r\\
    -\fr_l & 1\end{array}\right],
    && \det\bM=1,
    &&\bS=\left[\begin{array}{cc} \ft & \fr_r\\
    \fr_l & \ft
    \end{array}\right],
    \label{am-M=003}\\
    &\fr_l=-\frac{M_{21}}{M_{22}},
    &&\fr_r=\frac{M_{12}}{M_{22}}, &&\ft=\frac{1}{M_{22}}.
    \label{am-rt=M-2}
    \end{align}

Another consequence of (\ref{am-t=t}) is that the Wronskian of the Jost solutions take the form
    \be
    W(x)=\frac{2ik}{\ft(k)}.
    \label{am-wronskian}
    \ee
This is a number depending on the value of $k$. In particular, for $k\in\R^+$ it cannot diverge. This proves the following theorem.
    \begin{theorem}
    \label{am-thm2}
    Let $v(x)$ be a real or complex scattering potential. Then its transmission amplitude
    does not vanish for any wavenumber, i.e.,
    \be
    \ft(k)\neq 0~~~{\rm for}~~~k\in\R^+.
    \label{am-t=notzero}
    \ee
        \end{theorem}
This theorem shows that real and complex scattering potentials can never serve as a perfect absorber. According to Theorem~\ref{am-thm1} they cannot even serve as an approximate one-way filter.

Next, we examine the following simple example:
    \be
    v(x)=\fz\bigchi_{[0,L]}(x)=\left\{\begin{array}{ccc}
    \fz & {\rm for} & x\in[0,L],\\
    0 & {\rm for} & x\notin[0,L],
    \end{array}\right.
    \label{am-barrier}
    \ee
where $\fz$ and $L$ are respectively nonzero complex and real
parameters. This  is a piecewise constant finite-range
potential with support $[0,L]$, which we can identify with a rectangular
barrier potential of a possibly complex height $\fz$.

We can easily solve the Schr\"odinger equation~(\ref{am-sch-eq}) for
the barrier potential~(\ref{am-barrier}). Its general solution has
the form
    \be
    \psi(x)=\left\{\begin{array}{ccc}
    A_-(k)e^{ikx}+B_-(k)e^{-ikx} & {\rm for} & x<0,\\
    A_0(k)e^{i k\fn x}+B_0(k)e^{-ik\fn x}& {\rm for} & x\in[0,L],\\
    A_+(k)e^{ikx}+B_+(k)e^{-ikx} & {\rm for} & x\geq L,
    \end{array}\right.
    \label{am-barrier-sol-gen}
    \ee
where $A_j(k)$ and $B_j(k)$, with $j=0,\pm$, are  complex-valued
coefficient functions,
    \be
    \fn:=\sqrt{1-\frac{\fz}{k^2}},
    \label{am-ref-index=}
    \ee
and for every complex number $w$ we use $\sqrt w$ to  label the
principal value of $w^{1/2}$, i.e., $\sqrt w=\sqrt{|w|}e^{i\varphi}$
with $\varphi\in[0,\pi)$. By demanding $\psi$ to be continuous and
differentiable at $x=L$ and $x=0$, we can respectively express
$A_{+}$ and $B_{+}$ in terms of $A_0$ and $B_0$, and $A_{0}$ and
$B_{0}$ in terms of $A_-$ and $B_-$. This in turn allows us to
relate $A_{+}$ and $B_{+}$ to $A_-$ and $B_-$. We can write the
resulting equations in the form (\ref{am-M-def}) with the transfer matrix given
by
    \be
    \bM(k)=\left[\begin{array}{cc}
    [\cos(kL\fn)+i\fn_+\sin(kL\fn)]e^{-ikL} &
    i\fn_-\sin(kL\fn)e^{-ikL}\\[6pt]
    -i\fn_-\sin(kL\fn)e^{ikL} &
    [\cos(kL\fn)-i\fn_+\sin(kL\fn)]e^{ikL}\end{array}\right],
    \label{am-barrier-M}
    \ee
and $\fn_\pm:=(\fn\pm\fn^{-1})/2$.

In view of (\ref{am-rt=M-2}), we can use (\ref{am-barrier-M}) to read off the expression for the reflection and transmission amplitudes of the barrier potential~(\ref{am-barrier}). These have the form:
    \bea
    \fr_l(k)&=&\frac{i\fn_-\tan(kL\fn)}{1-i\fn_+\tan(kL\fn)},
    \label{am-barrier-r-left}\\
    \fr_r(k)&=&\frac{i\fn_-\tan(kL\fn)e^{-2ikL}}{1-i\fn_+\tan(kL\fn)},
    \label{am-barrier-r-right}\\
    \ft(k)&=&\frac{e^{-ikL}}{\cos(kL\fn)-i\fn_+\sin(kL\fn)}.
    \label{am-barrier-t}
    \eea
Clearly, $\ft(k)\neq 0$ for all $k\in\R^+$. We can check that indeed $\det\bM(k)=1$, and evaluate the $\bS$-matrix and its eigenvalues. In light of (\ref{am-S-eigenvalues}) the latter are given by
    \be
    \fs_\pm(k)=\left[\frac{1\pm i\fn_-\tan(kL\fn)}{1-i\fn_+\tan(kL\fn)}\right]e^{-ikL}.
    \label{am-barrier-S-eigenvalues}
    \ee

According to (\ref{am-barrier-r-left}) the barrier potential~(\ref{am-barrier}) is left-reflectionless if and only if $\fn$ is real and $k=k_m:=\pi m/L\fn$ for a positive integer $m$.\footnote{Equation~(\ref{am-ref-index=}) implies that $k_m=\sqrt{(\pi m/L)^2+\fz}$. This in turn means that for $\fz>0$, $m$ can be any positive integer, and for $\fz<0$, $m>L\sqrt{-\fz}/\pi$.} In this case it is also right-reflectionless, but not in general transparent. It is easy to show that for these values of the wavenumber, $\ft(k)=e^{-im\pi(\fn^{-1}+1)}$. This equals unity, i.e., the potential is transparent and hence bidirectionally invisible if and only if there is an integer $q$ such that $\fn=(2q/m-1)^{-1}$. It is easy to see that this is equivalent to demanding that
    \begin{align*}
    &\fz=\frac{4\pi^2q(q-m)}{L^2}, && k=\frac{2q-m}{L}.
    \end{align*}
Because $k>0$, the latter relation implies that $2q>m$.

The entries of the transfer matrix for the barrier potential~(\ref{am-barrier}) are smooth functions of the wavenumber $k$. In fact, we can analytically continue them to the entire complex $k$-plane. This turns out to be a common feature of all finite-range potentials. To see this first we note that if a potential $v(x)$ decays exponentially as $x\to\pm\infty$, i.e., there are positive numbers $\mu_\pm$ satisfying (\ref{am-exp-decay}), then the Jost solutions are holomorphic (complex analytic) functions in the strip \cite{am-blashchak-1968}:
    \be
    \sS_{\!\mu_\pm}:=\left\{ k\in\C~|~-\mu_-<\IM(k)<\mu_+ \right\}.
    \label{am-strip}
    \ee
In light of (\ref{am-jost2}) and the fact that $\det\bM=1$, this implies that the same holds for the entries of the transfer matrix. We state this result as a theorem:
    \begin{theorem}
    Let $v(x)$ be a real or complex potential satisfying (\ref{am-exp-decay}) for some $\mu_\pm>0$. Then the entries $M_{ij}(k)$ of its transfer matrix are holomorphic functions in the strip (\ref{am-strip}).
    \label{am-thm3}
    \end{theorem}

A basic result of complex analysis is that a nonzero holomorphic function can only vanish at a discrete set of isolated points. In view of Theorem~\ref{am-thm3} this applies to the entries of the transfer matrix of exponentially decaying potentials. In particular, for each choice of $i$ and $j$ in $\{1,2\}$, either $M_{ij}(k)=0$ for all $k\in\sS_{\!\mu_\pm}$ or there is a (possibly empty) discrete set of isolated values of $k\in\sS_{\!\mu_\pm}$ at which $M_{ij}(k)$ vanishes. This is particularly important, because $\sS_{\!\mu_\pm}$ contains the positive real axis where the physical wavenumbers reside.

According to (\ref{am-rt=M-2}), the zeros of $M_{12}(k)$ (respectively $M_{21}(k)$) that are located on the positive real axis are the wavenumbers $k_0$ at which the right (respectively left) reflection amplitude of the potential $v(x)$ vanishes, i.e., $v(x)$ is right- (respectively left-) reflectionless at $k_0$. Similarly, if $M_{22}(k_0)=1$, then $\ft(k_0)=1$, and $v(x)$ is transparent at $k_0$. Therefore real and positive zeros of $M_{12}(k)$, $M_{21}(k)$, and $M_{22}(k)-1$ are the wavenumbers at which $v(x)$ is right-reflectionless, left-reflectionless, and transparent. In particular, equations
    \bea
    &&M_{12}(k)=M_{22}(k)-1=0,
    \label{am-right-invisible}\\
    &&M_{21}(k)=M_{22}(k)-1=0,
    \label{am-leftt-invisible}
    \eea
respectively characterize the invisibility of the potential from the right and left. These results are clearly valid for any scattering system whose scattering features can be described using a transfer matrix.

The following no-go theorem is a simple consequence of Equations~(\ref{am-rt=M-2}) and the above-mentioned property of the zeros of holomorphic functions.
    \begin{theorem}
    If the entries $M_{ij}(k)$ of the transfer matrix for a scattering system are nonzero functions that are holomorphic on the positive real axis in the complex $k$-plane, then the system cannot display broadband reflectionlessness, transparency, or invisibility from either direction.
    \label{am-thm4}
    \end{theorem}
According to Theorem \ref{am-thm3}, the conclusion of this theorem applies to exponentially decaying and finite-range potentials.

The above analysis does not exclude the existence of exponentially decaying potentials that are unidirectionally or bidirectionally reflectionless for all $k\in\R^+$ (fullband reflectionlessness). Such potentials were known to exist since the 1930's. The principal example is the P\"oschl-Teller potential:
    \[v(x)=-\frac{\zeta}{\cosh(\alpha x)},\]
where $\zeta$ and $\alpha$ are positive real parameters. It turns out that the scattering problem for this potential admits an exact solution, and that for integer values of $\zeta/\alpha^2$ it is bidirectionally reflectionless for all $k\in\R^+$, \cite{am-flugge}. The P\"oschl-Teller potential is a member of an infinite class of real, attractive (negative), exponentially decaying potentials with this property. These were initially obtained in the 1950's as an application of the methods of inverse scattering theory \cite{am-kay-moses}. Their much less-known complex analogs were constructed in the 1990's, \cite{am-vu}.\footnote{Reflectionless potentials also arise as soliton solutions of nonlinear differential equations \cite{lamb}.}

The construction of scattering potentials that are unidirectionally invisible in the entire spectral band is a much more recent development \cite{am-horsley,am-longhi-2015}. The result is a class of complex potentials with  a power-law asymptotic decay behaviour. Before making specific comments on these potentials, we wish to address the problem of the existence of exponentially decaying and finite-range potentials that are unidirectionally reflectionless, transparent, or invisible in the whole spectral band. To do this, first we examine the structure of the transfer matrix $\bM(k)$ for negative values of $k$.

Consider a solution of the Schr\"odinger equation~(\ref{am-sch-eq}) for a scattering potential $v(x)$. In order to make the $k$-dependence of this solution explicit, we denote it by $\psi(k,x)$. In particular, we write (\ref{am-pw}) as
    \be
     \psi(k,x)\to A_\pm(k)e^{ikx}+B_\pm(k)e^{-ikx}\quad\quad {\rm for} \quad\quad
        x\to\pm\infty.
        \label{am-asymp-psi-k}
        \ee
Because the Schr\"odinger equation~(\ref{am-sch-eq}) is invariant under $k\to-k$,
    \be
    \breve\psi(k,x):=\psi(-k,x)
    \label{am-psi-breve}
    \ee
is also a solution of (\ref{am-sch-eq}).  In view of the fact that $v(x)$ is a scattering potential, $\breve\psi(k,x)$ must satisfy the asymptotic boundary conditions:
    \bea
    \breve\psi(k,x)&\to& \breve A_\pm(k)e^{-ikx}+\breve B_\pm(k)e^{ikx}\quad\quad
    {\rm for} \quad\quad x\to\pm\infty,
    \label{am-asymp-brevepsi-k}
    \eea
where $\breve A_\pm(k)$ and $\breve B_\pm(k)$ are some coefficient functions. We can use  (\ref{am-asymp-psi-k}), (\ref{am-psi-breve}), and (\ref{am-asymp-brevepsi-k}) to show that
for $k\in\R^-$,
    \begin{align}
    &\breve A_\pm(k)=B_\pm(-k),
    &&\breve B_\pm(k)=A_\pm(-k).
    \label{am-AB-k-reflection}
    \end{align}

Now, suppose that we can analytically continue $\bM(k)$ from $k\in\R^+$ to $k\in\R^-$. Then we can relate $\breve A_+(k)$ and $\breve B_+(k)$ to $\breve A_-(k)$ and $\breve B_-(k)$ using $\bM(k)$ for $k\in\R^-$. This gives
    \be
     \left[\begin{array}{c}
    \breve A_+(k)\\
    \breve B_+(k)\end{array}\right]=\bM(k)\left[\begin{array}{c}
    \breve A_-(k)\\
    \breve B_-(k)\end{array}\right].
    \label{am-M-breve}
    \ee
Substituting (\ref{am-AB-k-reflection}) in this equation and using (\ref{am-M-def}), we arrive at
    \be
    \bM(k)=\bsigma_1\bM(-k)\bsigma_1,
    \label{am-k-tomk}
    \ee
where $k\in\R^-$. Because this equation is invariant under $k\to-k$, it holds for all $k\in\R\setminus\{0\}$. In terms of the components of $\bM(k)$, we can write (\ref{am-k-tomk}) in the form:
    \begin{align}
    &M_{11}(-k)=M_{22}(k),
    &&M_{12}(-k)=M_{21}(k),
    \label{am-breve-Mij}
    \end{align}
which again hold for all  $k\in\R\setminus\{0\}$.

Equations (\ref{am-k-tomk}) and (\ref{am-breve-Mij}) apply to any scattering system in which the wave equation involves even powers of $k$ and have a transfer matrix that can be analytically continued from the positive to the negative real axis in the complex $k$-plane. For such systems,
we can determine the reflection and transmission amplitudes for $k\in\R^-$, by inserting (\ref{am-breve-Mij}) in (\ref{am-rt=M}). This gives
    \begin{align}
    & \fr_l(-k)=-\frac{\fr_r(k)}{\fD(k)},
    &&\ft_l(-k)=\frac{\ft_l(k)}{\fD(k)},
    &&\fr_r(-k)=-\frac{\fr_l(k)}{\fD(k)},
    &&\ft_r(-k)=\frac{\ft_r(k)}{\fD(k)},
    \label{am-btrvr-rt}
    \end{align}
where
    \be
    \fD(k):=\frac{M_{11}(k)}{M_{22}(k)}=\ft_l(k)\ft_r(k)-\fr_l(k)\fr_r(k)=\det\bS(k).
    \label{am-fD-def}
    \ee
Again, because Equations~(\ref{am-btrvr-rt}) are invariant under
$k\to-k$, they hold for all $k\in\R\setminus\{0\}$. A
straightforward consequence of these equations is that if
$\fr_{l/r}(k)$ (respectively $\ft_{l/r}(k)$) vanishes for all
$k\in\R^+$, then it will also vanish for all $k\in\R^-$. It is
important to note that this conclusion relies on the existence of
the analytic continuation of $\bM_{ij}(k)$ from $k\in\R^+$ to
$k\in\R^-$. Certainly, this condition holds for finite-range and
exponentially decaying potentials. This together with
Theorem~\ref{am-thm4} prove the following result.
    \begin{theorem}
    Scattering potentials with a finite range or an asymptotic exponential decay cannot display
    broadband unidirectional reflectionlessness, transparency, or invisibility.
    \label{am-thm-broadband}
    \end{theorem}
This theorem shows that as far as finite-range and exponentially decaying potentials are concerned, unidirectional reflectionlessness, transparency, and invisibility can only be achieved at a discrete set of isolated values of the wavenumber.

The principal example of a unidirectionally invisible finite-range potential is
    \be
    v(x)=\left\{\begin{array}{ccc}
    \fz\, e^{iK x} &{\rm for}&x\in[-\frac{L}{2},\frac{L}{2}],\\[3pt]
    0&{\rm for}&x\notin[-\frac{L}{2},\frac{L}{2}],
    \end{array}\right.
    \label{am-exp-pot}
    \ee
where $\fz$, $K$, and $L$ are respectively nonzero real parameters, and $L>0$, \cite{am-poladian,am-greenberg,am-kulishov,am-lin-2011}. This potential is unidirectionally invisible from the left for the wavenumber $k=K/2$, if $K=2\pi/L$ and $K^2\fz\ll 1$. It belongs to the class of locally periodic finite-range potentials of the form
    \be
    v(x)=\left\{\begin{array}{ccc}
    f(x) &{\rm for}&x\in[-\frac{L}{2},\frac{L}{2}],\\[3pt]
    0&{\rm for}&x\notin[-\frac{L}{2},\frac{L}{2}],
    \end{array}\right.
    \label{am-locally-periodic}
    \ee
where
    \be
    f(x):=\sum_{n=-\infty}^\infty \fz_n e^{iK_n x},
    \label{am-vn=}
    \ee
$\fz_n$ are complex coefficients, and $K_n:=2\pi n/L$. The following theorem, which is proven in Ref.~\cite{am-pra-2014a}, reveals a remarkable property of these potentials.
    \begin{theorem}
    Let $v(x)$ be a potential of the form (\ref{am-locally-periodic}) and suppose that we are interested in the scattering of waves of wavenumber $k$ satisfying $|\fz_n|/k^2\ll 1$, so that the first Born approximation is valid. If $\fz_n=0$ for all $n\leq 0$, $v(x)$ is unidirectionally left-invisible for all 
    $k=K_n/2=\pi n/L$.\footnote{This means that $v(x)$ is unidirectionally left-invisible for $k=K_n/2=\pi n/L$ 	     provided that we can neglect terms of order $(\fz_n/k^2)^2$ in the calculation of the reflection and transmission amplitudes.}
    \label{am-thm-invisibility}
    \end{theorem}

Now, consider taking $L\to\infty$. Then  (\ref{am-locally-periodic}) becomes $v(x)=f(x)$, the Fourier series in (\ref{am-vn=}) turns into a Fourier integral, the role of $\fz_n$ is played by the Fourier transform of $v(x)$, i.e., $\tilde v(\fK):=\int_{-\infty}^\infty e^{-i\fK x}v(x)dx$, and Theorem~\ref{am-thm-invisibility} states that {\em if the first Born approximation is reliable, then $v(x)$ is unidirectionally left-invisible for all $k\in\R^+$ provided that $\tilde v(\fK)=0$ for $\fK\leq 0$.} A highly nontrivial observation is that the same conclusion may be reached without assuming the validity of the first Born approximation \cite{am-horsley,am-longhi-2015}. In other words the following theorem on broadband invisibility holds.
    \begin{theorem}
    A scattering potential $v(x)$ is unidirectionally left-invisible for all wavenumbers $k\in\R^+$, if its Fourier transform $\tilde v(\fK)$ vanishes for all $\fK\leq 0$.
    \label{am-thm-horsley}
    \end{theorem}
Because the hypothesis of this theorem is equivalent to the condition that the real and imaginary part of $v(x)$ are connected by the spatial Kramers-Kronig relations, these potentials are sometimes called Kramers-Kronig potentials.\footnote{For a review of basic properties of these potentials, see \cite{am-horsley-longhi}.} It is well-known that they have a power-law decay at spatial infinities.\footnote{This explains why Theorems~\ref{am-thm-broadband} and \ref{am-thm-horsley} do not conflict.}

The unidirectional invisibility of the potential (\ref{am-exp-pot}) for $k=K/2=\pi/L$ is a perturbative result \cite{am-pra-2014a}; it is violated for sufficiently large values of $|\fz|$, \cite{am-longhi-JPA-2011,am-jones-2012}. This potential does however support exact (nonperturbative) unidirectional invisibility for particular values of $\fz$, \cite{am-jpa-2016}. Another example of a finite-range potential with exact unidirectional invisibility is (\ref{am-locally-periodic}) with
    \[f(x):=\frac{-2\alpha K^2(3-2 e^{i Kx})}{e^{2iKx}+\alpha(1-e^{iKx})^2},\]
where $\alpha$ and $K$ are real parameters. It turns out that this potential is unidirectionally right-invisible for $k=K/2=\pi n/L$ with $n$ being any positive integer provided that $\alpha>-1/4$, \cite{am-pra-2014b}. The simplest scattering potential supporting exact unidirectional invisibility are barrier potentials of the form $v(x)=\fz_1\chi_{[-a_1,0)}+\fz_2\chi_{[0,a_2]}$ where $\fz_j$ and $a_j$ are respectively complex and positive real parameters \cite{am-pra-2013}. See also \cite{mustafa-2017}.

\section{Spectral singularities, resonances, and bound states}
\label{am-SS}

In Section~\ref{am-S4} we show that the Wronskian of the Jost solutions $\psi_\pm$ of the Schr\"odinger equation for a scattering potential $v(x)$ is given by
    \be
    W(x)=\frac{2ik}{\ft(k)}=2ik M_{22}(k).
    \label{am-wronskian}
    \ee
This in particular implies that $\psi_\pm$ are linearly dependent solutions of the Schr\"odinger equation~(\ref{am-sch-eq}) whenever $k$ is a real and positive zero of $M_{22}(k)$. This represents a physical wavenumber $k$ at which $\ft(k)$ blows up. The corresponding value of the energy, $E:=k^2$, which belongs to the continuous spectrum of the Schr\"odinger operator, $-\frac{d^2}{dx^2}+v(x)$, is called a {\em spectral singularity}\footnote{The notion of a spectral singularity was originally introduced in \cite{am-Naimark} for Schr\"odinger operators in the half-line. It was subsequently generalized to the case of full-line in \cite{am-kemp-1958}. The term ``spectral singularity'' was originally used to refer to this notion in \cite{am-schwartz}. For a readable account of basic mathematical facts about spectral singularities and further references, see \cite{am-ghuseynov}.} of the potential \cite{am-prl-2009}.

If $k_0^2$ is a spectral singularity, $M_{22}(k_0)=0$, but because $\det\bM(k_0)=1$, neither of  $M_{12}(k_0)$ and $M_{21}(k_0)$ can vanish. In light of (\ref{am-rt=M-2}), this implies that similarly to the transmission amplitude $\ft(k)$, the reflection amplitudes $\fr_{l/r}(k)$ blow up at $k=k_0$. Furthermore, (\ref{am-jost2}) shows that whenever $M_{22}(k)=0$,
    \be
    \psi_+(x)=-\frac{M_{21}(k)}{M_{12}(k)}\,\psi_-(x)\to
    \left\{\begin{array}{ccc}
    -M_{21}(k)e^{-ikx} & {\rm for} & x\to-\infty,\\
    e^{ikx} & {\rm for} & x\to+\infty.\end{array}\right.
    \label{am-outgoing}
    \ee
Application of this relation for $k=k_0$ shows that at a spectral singularity Jost solutions $\psi_\pm(x)$ are scattering solutions of the Schr\"odinger equation that satisfy outgoing asymptotic boundary conditions. These are also known as the Seigert boundary conditions \cite{am-seigert-1939} which provide a standard description of resonances.

Consider a solution $\psi(x)$ of the time-independent Schr\"odinger equation~(\ref{am-sch-eq}) for a general complex value of the energy $k^2$ and suppose that it satisfies the outgoing asymptotic boundary conditions:
    \be
    \psi(x)\to N_\pm(k)\, e^{\pm ikx}~~{\rm for}~~x\to\pm\infty,
    \label{am-resonance}
    \ee
where $N_\pm(k)$ are nonzero complex coefficients. $\psi(x)$ corresponds to a solution $\psi(x,t)$ of the time-dependent Schr\"odinger equation, $i\partial_t\psi(x,t)=-\partial_x^2\psi(x,t)+v(x)\psi(x,t)$, namely
    \be
    \psi(x,t):=e^{-ik^2 t}\psi(x)=e^{-\Gamma t}e^{-iEt}\psi(x),
    \label{am-res-sol}
    \ee
where
    \begin{align}
    E:=\RE(k)^2-\IM(k)^2, && \Gamma:=-2\RE(k)\IM(k).
    \label{am-E-Gamma}
    \end{align}
If $\Gamma>0$, $\psi(x,t)$ decays exponentially as $t\to\infty$. In this case, we identify $\psi(x,t)$ with a {\em resonance}. The quantity $\Gamma$ which determines its decay rate is called the {\em width} of the resonance. If $\Gamma<0$, $\psi(x,t)$ grows exponentially as $t\to\infty$, and we call it an {\em antiresonance}. It is not difficult to see that resonances and antiresonances are also zeros of $M_{22}(k)$. But
the corresponding value of $k^2$ lie in the lower and upper complex energy half-planes,
    \begin{align*}
    &\cE_{\rm lower}:=\{ k^2\in\C~|~\IM(k^2)<0\},
    &&\cE_{\rm upper}:=\{ k^2\in\C~|~\IM(k^2)>0\},
    \end{align*}
respectively.

The Jost solutions of the time-independent Schr\"odinger equation~(\ref{am-sch-eq}) that correspond to a spectral singularity satisfy the above description of a resonance except that for a spectral singularity $k$ is real. This suggests identifying these solutions with certain {\em zero-width resonances} \cite{am-prl-2009}.\footnote{Spectral singularities must be distinguished with the solutions of the time-independent Schr\"odinger equation that correspond to a bound state in the continuum \cite{am-stillinger-1975,am-BSC} for the following reasons: 1) They define scattering states that do not decay at spatial infinities. 2) They may exist for exponentially decaying and short-range potentials. 3) As we explain in Section~\ref{am-Sec8}, real potentials cannot have spectral singularities. None of these holds for bound states in the continuum.} Note that spectral singularities lie on the positive real axis in the complex energy plane:
    \be
    \cE_+:=\{ k^2\in\C~|~\RE(k^2)>0~{\rm and}~\IM(k^2)=0\}.
    \label{am-SS-E-plane}
    \ee

There is another way in which we can have a real zero of $M_{22}(k)$ such that $\Gamma=0$. This is when $k$ is purely imaginary; i.e., $E=k^2\in\R^-$. Let us set $k=i\sqrt{|E|}$. Then, according to (\ref{am-outgoing}), $\psi_+$ determines a solution of the time-independent Schr\"odinger equation that decays exponentially at spatial infinities. This solution is clearly square-integrable. Therefore its energy $E=k^2$, which is real and negative, belongs to the point spectrum of the Schr\"odinger operator $-\frac{d^2}{dx^2}+v(x)$; it is a real and negative eigenvalue of this operator that corresponds to a bound state of the potential $v(x)$. If $k$ is a zero of $M_{22}(k)$ that lies in the upper-half $k$-plane, i.e., $\IM(k)>0$, then $|\psi_+(x)|$ is again exponentially decaying as $x\to\pm\infty$. Therefore $\psi_+(x)$ is  a square-integrable function and $k^2$ is a complex eigenvalue of $-\frac{d^2}{dx^2}+v(x)$.

Note that the above discussion of the interpretation of the zeros of $M_{22}(k)$ as spectral singularities, resonances, antiresonances, and eigenvalues of the Schr\"odinger operator $-\frac{d^2}{dx^2}+v(x)$ applies to any scattering potential. As shown in \cite{am-kemp-1958}, in this case the Jost solutions $\psi_\pm$ and consequently the entries of the transfer matrix are continuous functions of $k$ for $\IM(k)\geq 0$. They might not however be holomorphic in any region containing the real axis in the complex $k$-plane. If there is such a region in which $M_{22}(k)$ is a nonzero holomorphic function, then the zeros of $M_{22}(k)$ that lie in this region form a discrete isolated set of points. This in turn implies that one cannot have spectral singularities in an extended interval of real numbers other than the whole positive real axis. In particular we have the following result.
    \begin{theorem}
    If $v(x)$ is a real or complex potential with a finite range or an asymptotic exponential decay, so that (\ref{am-exp-decay}) holds for some $\mu_\pm\in\R^+$, then either its spectral singularities are isolated points of the positive real axis in the complex energy plane or cover the whole positive real axis.
    \label{am-thm5}
    \end{theorem}

Next, we examine the behavior of the eigenvalues $\fs_\pm$ of the $\bS$-matrix in the vicinity of a spectral singularity $k_0^2$. As $k\to k_0$,  $\epsilon:=M_{22}(k)$ tends to zero.  Because the entries of the transfer matrix are continuous functions on the upper half-plane and $\IM(k_0)\geq 0$, none of them blow up at $k=k_0$. We also know that $\det\bM(k)=1$. In view of these observations and (\ref{am-S-eigenvalues}), we can show that the eigenvalues of the $\bS$-matrix for a scattering potential satisfy
    \be
    \fs_\pm(k)\to\frac{1}{\epsilon}\pm\frac{1}{|\epsilon|}\mp
    \frac{{\rm sgn}(\epsilon)M_{11}(k_0)}{2}\quad\quad{\rm for}\quad\quad k\to k_0.
    \label{am-S-eigenvalue-ss}
    \ee
This implies that as $k^2$ approaches a spectral singularity, one of the eigenvalues of $\bS(k)$ diverges while the other attains a finite limit. More specifically we have the following result.
    \begin{theorem}
    Let $k_0^2$ be a spectral singularity of a scattering potential $v(x)$. Then as $k\to k_0$ the eigenvalues (\ref{am-S-eigenvalues}) of the $\bS$-matrix behave as follows. Either $\fs_-(k)\to- M_{11}(k_0)/2$ and $|\fs_+(k)|\to\infty$, or
$|\fs_-(k)|\to\infty$ and $\fs_+(k)\to  M_{11}(k_0)/2$.
    \label{am-thm6}
    \end{theorem}

Now, suppose that $v(x)$ is a scattering potential such that  $\det\bS(k)$ is a bounded function of $k$. Then
Theorem~\ref {am-thm6} implies that $M_{11}(k_0)=0$ whenever $k_0^2$ is a spectral singularities of $v(k)$, i.e., $k_0$ is a common zero of $M_{11}(k)$ and $M_{22}(k)$. Spectral singularities satisfying this condition are said to be {\em self-dual} \cite{am-jpa-2012}. We study these in Section~\ref{am-Sec9}.

Let us examine the spectral singularities of a couple of exactly solvable potentials.

First, consider a delta-function potential with a complex coupling constant $\fz$,  \cite{am-jpa-2006},
    \be
    v(x)=\fz\,\delta(x).
    \label{am-delta-potential}
    \ee
We can determine its transfer matrix using (\ref{am-pt-int-delta}), (\ref{am-pt-int-N=}), and (\ref{am-pt-int-M=}) with $c=0$. This gives
    \be
    \bM(k)=\left[\begin{array}{cc}
    1-i\fz/2k & -i\fz/2k\\
    i\fz/2k & 1+i\fz/2k\end{array}\right].
    \label{am-M-delta}
    \ee
In view of this relation and (\ref{am-rt=M-2}),
    \begin{align}
    &\fr_l(k)=\fr_r(k)=\frac{-i\fz}{2k+i\fz},
    &&\ft(k)=\frac{2k}{2k+i\fz}.
    \label{am-r-t-delta}
    \end{align}

The following are consequences of the fact that $M_{22}(k)$ has a single zero, namely $k_0=-i\fz/2$.
    \begin{itemize}
    \item The delta-function potential has a spectral singularity, if and only if  $\fz$ is purely imaginary and $\IM(\fz)>0$, i.e., $\fz=i\zeta$ for some $\zeta\in\R^+$. In this case, $k_0=\zeta/2$, the spectral singularity has the value $k_0^2=\zeta^2/4$, and
        \be
        \psi_+(x)=e^{\pm ik_0x}\quad\quad {\rm for} \quad\quad  \pm x\geq 0.
        \label{am-delta-jost}
        \ee
    \item It has a single resonance (respectively antiresonance) with a square-integrable position wave function $\psi(x)$ if and only if $\IM(\fz)>0$ (respectively $<0$) and $\RE(\fz)<0$. In this case $\psi(x)$ is a constant multiple of the right-hand side of (\ref{am-delta-jost}) with $k_0=
    [\IM(\fz)-i\RE(\fz)]/2$.
    \item It has a bound state with a real and negative energy if and only if $\fz\in\R^-$. The position wave function for this state is a constant multiple of the right-hand side of (\ref{am-delta-jost}) with $k_0=i|\fz|/2$.
    \end{itemize}

Next, we consider the spectral singularities of the complex barrier potential~(\ref{am-barrier}). According to (\ref{am-barrier-M}), zeros $k_0$ of $M_{22}(k)$ satisfy
    \be
    \cos(k_0L\fn_0)-i\fn_{0+}\sin(k_0L\fn_0)=0,
    \label{am-barrier-ss-0}
    \ee
where
    \begin{align}
    &\fn_0:=\sqrt{1-\frac{\fz}{k_0^2}}, &&\fn_{0+}:=\frac{\fn_0^2+1}{2\fn_0}.
    \end{align}
It is not difficult to express (\ref{am-barrier-ss-0}) in the form:
    \be
    e^{-2ik_0L\fn_0}=\left(\frac{\fn_0-1}{\fn_0+1}\right)^2,
    \label{am-barrier-ss-1}
    \ee
$k_0^2$ is a spectral singularity if and only if $k_0$ is a positive real number satisfying this relation. For such a $k_0$, we can write (\ref{am-barrier-ss-1}) as a pair of real equations for the $k_0$, $\eta_0:=\RE(\fn_0)$, and $\kappa_0:=\IM(\fn_0)$. Because
    \be
    \fn_0=\eta_0+i\kappa_0,
    \label{am-fn-zero}
    \ee
evaluating the modulus of both side of (\ref{am-barrier-ss-1}) we find
    \be
    \kappa_0=\frac{1}{2k_0L}\ln\left|\frac{(\eta_0-1)^2+\kappa_0^2}{
    (\eta_0+1)^2+\kappa_0^2}\right|.
    \label{am-kappa-zero}
    \ee
Similarly, equating the phase angles of both side of (\ref{am-barrier-ss-1}), we obtain
    \be
    k_0=\frac{2\pi m-\varphi_0}{2L\eta_0},
    \label{am-k-zero}
    \ee
where $m$ is a positive integer, and $\varphi_0$ is the principle argument of the right-hand side of (\ref{am-barrier-ss-1}), i.e.,
    \begin{align}
    &\varphi_0=\left\{\begin{array}{ccc}
    \arctan(\alpha_0)& {\rm for} &
    \eta_0^2+\kappa_0^2\geq 1,\\
    \arctan(\alpha_0)-\pi & {\rm for} &
    \eta_0^2+\kappa_0^2< 1,\end{array}\right.
    && \alpha_0:=\frac{2\kappa_0}{(\eta_0^2+1)^2+\kappa_0^2}.
    \end{align}

Next, let us identify the barrier potential~(\ref{am-barrier}) with an optical potential (\ref{am-opt-pot}) that describes the scattering of normally incident polarized electromagnetic waves by an infinite slab of homogeneous nonmagnetic material. We choose a coordinate system in which the slab occupies the space confined between the planes $x=0$ and $x=L$, and the wave is polarized along the $y$-direction and propagates along the $x$-direction. Then the relative permittivity of the system that enters the Helmholtz equation~(\ref{am-helm-eq}) has the form:
    \be
    \hat\varepsilon(x)=\left\{\begin{array}{ccc}
    \hat\varepsilon_{\rm slab} & {\rm for} & x\in[0,L],\\
    1  & {\rm for} & x\notin[0,L],\end{array}\right.
    \label{am-varepsilon=}
    \ee
where $\hat\varepsilon_{\rm slab}$ is the relative permittivity of the slab. In general this takes a possibly complex constant value. We can identify the Helmholtz equation with the Schr\"odinger equation~(\ref{am-sch-eq}) provided that $v(x)$ is the barrier potential~(\ref{am-barrier}) with $\fz=k^2(1-\hat\varepsilon_{\rm slab})$. Substituting this equation in (\ref{am-ref-index=}), we find $\fn=\sqrt{\hat\epsilon}$. Therefore $\fn$ is the refractive index of the slab.

According to (\ref{am-kappa-zero}) the optical system we have described has a spectral singularity, if the imaginary part of the refractive index of our slab is negative. This is precisely the case where the slab is made out of gain material. To see this we note that the gain coefficient of a homogeneous medium is related to its refractive index according to
    \be
    g=-\frac{4\pi\IM(\fn)}{\lambda}=-2k\IM(\fn),
    \label{am-gain}
    \ee
where $\lambda=2\pi/k$ is the wavelength \cite{am-silfvast}. If the refractive index of the slab equals $\fn_0$, it emits coherent outgoing radiation of wavelength $\lambda_0=2\pi/k_0$, i.e., it acts as a laser. In view of (\ref{am-kappa-zero}), for $k=k_0$ and $\fn=\fn_0$, the gain coefficient (\ref{am-gain}) is given by \cite{am-pra-2011a}:
    \be
    g=\frac{1}{L}\ln\left|\frac{(\eta_0+1)^2+\kappa_0^2}{
    (\eta_0-1)^2+\kappa_0^2}\right|=\frac{2}{L}
    \ln\left|\frac{\fn_0+1}{\fn_0-1}\right|.
    \label{am-g-zero}
    \ee
This relation is known as the {\em laser threshold condition}  in
optics \cite{am-silfvast}. It is usually derived by balancing the
energy input of the laser by the sum of its energy output and losses. Here we
obtain it using the notion of spectral singularity, i.e., demanding
the existence of purely outgoing solutions of the wave equation.
Notice that this condition also yields a formula for the available
laser modes, namely (\ref{am-k-zero}). For typical lasers, $k_0L\gg
1$. This implies $m\gg1$ which together with (\ref{am-k-zero}) give
$k_0\approx \pi m/L\RE(\fn_0)$. The latter is also a well-known
result in optics.

The notion of spectral singularity can be extended to more general
scattering problems. This is done by identifying it with the values
of $k^2$ at which the left or right reflection and transmission
coefficients blow up.  This corresponds to situations where
$\psi(x)$ satisfies purely outgoing boundary
conditions.\footnote{The importance of purely outgoing waves in the
laser theory predates the discovery of their connection to the
mathematics of spectral singularities. See for example
\cite{am-tureci-2006}.} For a linear scattering problem,  the
assumption $\det\bM(k)\neq 0$ together with Equations
(\ref{am-rt=M}) imply that spectral singularities are given by the
real and positive zeros of $M_{22}(k)$ and that they are always
bidirectional, i.e., both the left and right reflection and
transmission coefficients diverge at a spectral singularity.

Determination of spectral singularities of an optical system having an arbitrary geometry is equivalent to finding its laser threshold condition. This observation has been employed for obtaining laser threshold condition for bilayer \cite{am-jpa-2012}, cylindrical \cite{am-pra-2013b}, and spherical \cite{am-pla-2011,am-prsa-2012,am-pra-2013d} lasers. A brief review of the physical aspects of spectral singularities is provided in \cite{am-SS-review}. For a discussion of the spectral singularities of nonlinear Schr\"odinger equation and their applications in optics, see \cite{am-prl-2013,am-gupta-2014,am-jo-2017,p136}.

\section{Space-reflections and time-reversal transformation}
\label{am-S6}

In this section we explore the space-reflection and time-reversal transformations in quantum mechanics. This requires the knowledge of unitary and Hermitian operators acting in a Hilbert space. Because a precise definition of a Hermitian operator involves certain notions of functional analysis that are not familiar to most physicists, here we provide a less rigorous description of this concept. The interested reader may consult \cite{am-review,prugovecki} for a more careful treatment of the subject. 

Consider a linear operator $L$ acting in a Hilbert space $\sH$, and let $\pbr\cdot\,,\cdot\pkt$ denote the inner product of $\sH$. Then the {\em adjoint} of $L$ is the operator $L^\dagger:\sH\to\sH$ that satisfies
    \[\pbr\cdot, L\,\cdot\pkt = \pbr L^\dagger \cdot\,,\cdot\pkt.\]
We call $L$ {\em Hermitian} or {\em self-adjoint} if $L^\dagger=L$. We call it a {\em unitary operator} if its domain is $\sH$, it is one-to-one and onto, and $L^{-1}=L^\dagger$. These conditions are equivalent to the requirement that
    \[\pbr L\phi_1,L\phi_2\pkt=\pbr\phi_1,\phi_2\pkt,\]
i.e., $L$ leaves the inner product invariant. Here and in what follows $\phi_1$ and $\phi_2$ are arbitrary elements of $\sH$. It turns out that $L$ is unitary if and only if it preserve the norm of the vectors; $\parallel L\phi_1\parallel=\parallel\phi_1\parallel$ where $\parallel\phi_1\parallel:=\sqrt{\pbr\phi_1,\phi_1\pkt}$.

In the standard quantum mechanical description of the nonrelativistic motion of a
particle on a straight line, we take $\sH$ to be the space of square
integrable functions $L^2(\R)$ endowed with the inner product:
$\br\phi_1|\phi_2\kt:=\int_{-\infty}^\infty\phi_1(x)^*\phi_2(x)dx$.

Hermitian operators play a basic role in both kinematical and
dynamical aspects of quantum mechanics. Observables of quantum
systems are described by Hermitian operators not just because they
have a real spectrum, but more importantly because their expectation
values are real. {\em Non-Hermitian operators may have a real
spectrum and even a complete set of eigenvectors forming a basis of
the Hilbert space, but there are always states in which their
expectation value is not real.}\footnote{For a proof of this
statement see \cite[Appendix]{am-review}. A more detailed discussion
is provided in \cite{am-schechter}.} Because the calculation of
expectation values involves the inner product of the Hilbert space,
a non-Hermitian operator can play the role of an observable of a
quantum system, only if we can  modify the inner product on the
space of state vectors or even the space of state vectors itself
\cite{am-ptrsa-2013}, so that the operator acts in the new Hilbert
space as a Hermitian operator.\footnote{This is obviously not always
possible. A sufficient condition for the existence of such a
modified inner product is that the operator $L$ satisfies the
pseudo-Hermiticity relation $L^\dagger=\eta\, L\eta^{-1}$ for a
positive-definite bounded linear operator $\eta$ with a bounded
inverse. For operators acting in finite-dimensional Hilbert spaces
this is also a necessary condition. For further discussion of these
and related issues see \cite{am-review,am-ptrsa-2013} and references
therein.} This leads to different representations of quantum
mechanics whose structure is identical to the standard
representation that we employ here \cite{am-review}. The Hamiltonian
operator is required to be Hermitian not only because it is usually
identified with the energy observable, but also because it ensures
the unitarity of time-evolution, i.e., the time-evolution operator
defined by the Hamiltonian is a unitary operator. A celebrated
result of functional analysis, known as Stone's theorem
\cite{am-reed-simon}, establishes the converse of this statement.
Therefore, {\em the unitarity of dynamics implies the Hermiticity of
the Hamiltonian.} This result also disqualifies non-Hermitian
operators from serving as the Hamiltonian operator for a unitary
quantum system.

Non-Hermitian operators can nevertheless be employed in the study of open quantum systems and a variety of problems in the areas where some of the axioms of quantum mechanics are violated. This has actually turned out to be more fruitful than the attempts to use non-Hermitian operators for invoking the nonstandard representations of quantum mechanics.

Having reviewed the meaning of Hermiticity and unitarity of an
operator and their role in quantum mechanics, we return to the study
of space-reflections and time-reversal transformation.

For each $a\in\R$, the active transformation, $x\to 2a-x$,
corresponds  to the reflection of the real line about the point $a$.
This transformation induces a mapping of the wave functions
$\phi(x)$ according to $\phi(x)\to\phi(2a-x)$. We identify this with
the action of a linear operator $\cP_{a}$ in $L^2(\R)$, namely
$\phi\to\widetilde\phi:=\cP_a\phi$, where
    \be
    (\cP_{a}\phi)(x):=\phi(2a-x).
    \label{am-gen-parity}
    \ee
It is easy to show that $\cP_{a}$ is a Hermitian operator. It is
also  clear that $\cP_{a}^2=I$, so that $\cP_{a}^{-1}=\cP_{a}$.
Combining this with the Hermiticity of $\cP_{a}$ we conclude that
$\cP_{a}$ is also a unitary operator.

We can use $\cP_{a}$ to transform linear operators $L(t)$ acting in
$L^2(\R)$ according to
    \be
    L(t)\to \widetilde{L}(t):=\cP_{a}\,L(t)\,\cP_{a}^{-1}=\cP_{a}\,L(t)\,\cP_{a}.
    \label{am-L-parity-trans}
    \ee
For example, let $\hat x$, $\hat p$, and $H(t)$ be respectively the
standard position, momentum, and Hamiltonian operators acting in
$L^2(\R)$, i.e.,
    \begin{align}
    &\hat x\,\phi(x):=x\phi(x),
    &&\hat p\,\phi(x):=-i\phi'(x),
    &&H(t)=\frac{\hat p^2}{2m}+v(\hat x,t).
    \label{am-standard-t}
    \end{align}
We can use (\ref{am-gen-parity}) to show that
    \begin{align}
    &\{\hat x,\cP_a\}=2aI,
    &&\{\hat p,\cP_a\}=0,
    \label{am-comm-Pa-xp}
    \end{align}
where $\{\,\cdot\,,\,\cdot\,\}$ stands for the anticommutator of
operators. Equations (\ref{am-L-parity-trans}) --
(\ref{am-comm-Pa-xp}) imply
    \begin{align}
    &\widetilde{\hat x}=2aI-\hat x, &&
    \widetilde{\hat p}=-\hat p,
    && \widetilde{H}(t)=\frac{\hat p^2}{2m}+v(2a I-\hat x,t).
    \end{align}
The first of these relations justifies the name ``space-reflection''
or  ``parity-operator with respect to $a$'' for $\cP_{a}$.

If $H(t)$ is the Hamiltonian operator for a quantum system $\cS$, we call the quantum system defined by $\widetilde H$  the ``space-reflection of $\cS$ with respect to $a$.'' Equation (\ref{am-L-parity-trans}) and the unitarity of $\cP_a$ imply that $\widetilde H(t)$ is Hermitian if and only if so is $H(t)$. This means that space-reflections of a unitary quantum system are unitary.

An operator $L(t)$ is called  {\em parity-invariant with respect to
$a$} if $\widetilde{L}(t)=L(t)$. In particular, a standard
Hamiltonian operator (\ref{am-standard-t}) is parity-invariant with
respect to $a$ if and only if $v(2a -x,t)=v(x,t)$.

The parity operators $\cP_a$ can be  generated from $\cP_0$ using
the space-translation operator $T_{a}:=e^{-ia\hat p}$ which
satisfies:
    \be
    (T_{a}\,\phi)(x)=\phi(x-a).
    \label{am-space-trans}
    \ee
To see this we use (\ref{am-gen-parity}) and (\ref{am-space-trans})
to show that
    \[(\cP_a\phi)(x)=\phi(2a-x)=(\cP_0\phi)(x-2a)=(T_{2a}\cP_0\phi)(x).\]
Therefore,
    \be
    \cP_a=T_{2a}\cP_0.
    \label{am-parity=}
    \ee

We use the symbol $\cP$ for $\cP_0$ and refer to it as the {\em
parity operator} in $L^2(\R)$. According to this terminology a
standard Hamiltonian operator (\ref{am-standard-t}) is {\em
parity-invariant} or {\em $\cP$-symmetric} if and only if
$v(x,t)=v(-x,t)$. For a time-independent potential $v(x)$, this
means that it is an even function.

Next, consider the operation of complex-conjugation of
complex-valued functions, $\phi(x)\to\phi(x)^*$. This defines a
function $\cT:L^2(\R)\to L^2(\R)$ according to
$(\cT\phi)(x):=\phi(x)^*$. Because for any pair of complex numbers
$\alpha_1$ and $\alpha_2$,
    \[\cT(\alpha_1\phi_1+\alpha_2\phi_2)=\alpha_1^*\cT\phi_1+\alpha_2^*\cT\phi_2,\]
$\cT$ is an {\em antilinear operator}. It is also clear  that $\cT$
squares to the identity operator $I$. In particular, it is
invertible, and $\cT^{-1}=\cT$.

Let us apply $\cT$ to both sides of the time-dependent Schr\"odinger equation,
    \be
    i\frac{d}{dt}\psi(x,t)=H(t)\psi(x,t).
    \label{am-TD-Sch-eq-gen}
    \ee
This gives $-i\frac{d}{dt}\cT\psi(x,t)=\cT H(t)\psi(x,t)$. We can write
this equation in the form
    \be
    i\frac{d}{d(-t)}\cT\psi(x,t)=\overline H(-t)\cT\psi(x,t),
    \label{am-TD-Sch-rq5}
    \ee
where for a time-dependent linear operator $L(t)$,
    \be
    \overline L(t):=\cT L(-t)\cT^{-1}=\cT L(-t)\cT.
    \label{am-H-tide}
    \ee
If we make the change of variables:
    \begin{align*}
    &t\to \overline t:=-t,
    &&\psi(x,t)\to\overline\psi(x,\overline t):=\psi(x,t)^*=(\cT\psi)(x,t),
    \end{align*}
(\ref{am-TD-Sch-rq5}) takes the form  $i\frac{d}{d \overline
t}\overline\psi(x,\overline t)=\overline H(\overline t))
\overline\psi(x,\overline t)$. Because $t$ and $\overline t$ take
arbitrary real values, this equation is equivalent to
    \be
    i\frac{d}{d t}\overline \psi(x, t)=\overline H(t) \overline\psi(x, t).
    \label{am-cT-sch-eqn2}
    \ee

We can express the solutions of (\ref{am-TD-Sch-eq-gen})  and
(\ref{am-cT-sch-eqn2}) in terms of the time-evolution operators
$U(t)$ and $\overline U(t)$ for the Hamiltonians $H(t)$ and
$\overline H(t)$. For a given initial state vector $\psi_0(x)$, we
have
    \begin{align}
    &\psi(x,t)=U(t)\psi_0(x), &&\overline\psi(x,t)=\overline U(t)
    \psi_0(x)^*.
    \end{align}
According to these relations, as we increase the value of  the time
label $t$ starting from $t=0$, the evolution operators $U(t)$ and $\overline U(t)$
respectively determine $\psi(x,t)$ and $\overline\psi(x,t)$ for $t>0$. 
In view of the fact that $\psi(x,-t)=\overline\psi(x,t)^*$, we can say that 
$\overline U(t)$ determines $\psi(x,t)$ for $t<0$. For this reason, the systems
described by the Hamiltonian operators $H(t)$ and $\overline H(t)$
are said to be the time-reversal of one another. This, in
particular, suggests identifying the antilinear operator $\cT$ with the {\em
time-reversal operator}.

The above argument leaves a crucial question unanswered:  Suppose
that $H(t)$ is a Hermitian operator so that it determines a unitary
quantum system. Does this imply that the time-reversed system is
also unitary? Equivalently, is $\overline H(t)$ Hermitian? The
answer turns out to be in the affirmative, because $\cT$ satisfies
    \be
    \br\cT\phi_1|\cT\phi_2\kt=\int_{-\infty}^\infty [\cT\phi_1(x)]^*\cT\phi_2(x)dx=
    \int_{-\infty}^\infty \phi_1(x)\phi_2(x)^*dx=\br\phi_2|\phi_1\kt.
    \label{am-antiunitary}
    \ee
With the help of this relation and the Hermiticity of $H(t)$, we can show that
    \bea
    \br\phi_1|\overline H(t)\phi_2\kt&=&\br\cT^2\phi_1|\overline H(t)\phi_2\kt=
    \br \cT^2\phi_1|\cT H(-t)\cT\phi_2\kt
    =\br H(-t)\cT\phi_2|\cT\phi_1\kt\nn\\
    &=&\br\cT\phi_2|H(-t)\cT\phi_1\kt=\br\cT\phi_2|\cT^2H(-t)\cT\phi_1\kt=
    \br\cT H(-t)\cT\phi_1|\phi_2\kt\nn\\
    &=&\br\overline H(t)\phi_1|\phi_2\kt.\nn
    \eea
This concludes the proof of the Hermiticity of $\overline H(t)$.

An antilinear operator $\fS$, which by definition satisfies
    \[\fS(\alpha_1\phi_1+\alpha_2\phi_2)=
    \alpha_1^*\fS\phi_1+\alpha_2^*\fS\phi_2,\]
is said to be unitary, if
    \be
    \br\fS\phi_1|\fS\phi_2\kt=\br\phi_2|\phi_1\kt.
    \label{antiunitary}
    \ee
Unitary antilinear operators are also called ``{\em antiunitary operators}'' \cite{am-weinberg}.
Similarly to unitary linear operators they preserve the norm of state vectors.

Equation~(\ref{am-antiunitary}) means that $\cT$ is an antiunitary operator. There are other antiunitary operators that square to identity and share the time-reversal property of
$\cT$.\footnote{A simple examples is $\cT_\tau:=e^{i\tau}\cT$ where
$\tau\in\R$.} This implies that in general $\cT$ is not the only
possible choice for a time-reversal operator \cite{am-messiah}. In
what follows, however, we take $\cT$ to implement the time-reversal
transformation in $L^2(\R)$ and refer to it as the {\em
time-reversal operator}.

A possibly time-dependent linear operator $L(t)$ is  said to be {\em
time-reversal-invariat} or {\em real} if $\overline L(t)=L(t)$. It
is called an {\em imaginary operator} if $\overline L(t)=-L(t)$. For
example, the standard position operator $\hat x$ is real, because
    \[\overline{\hat x}\,\phi(x)=\cT \hat x \cT\phi(x)=[x\phi(x)^*]^*=x\phi(x)=\hat x\,\phi(x),\]
while the standard momentum operator $\hat p$ is imaginary, because
    \[\overline{\hat p}\,\phi(x)=\cT \hat p \cT\phi(x)=
    \cT\left[-i\frac{d}{dx}\phi(x)^*\right]=
    i\cT\left[\frac{d}{dx}\phi(x)^*\right]=i\frac{d}{dx}\phi(x)=-\hat p\phi(x).\]
Clearly $L(t)$ is an imaginary operator if and  only if $iL(x)$ is
real. In particular, $iI$ is imaginary, because $I$ is a real
operator. Note also that time-independent linear operators $L_R$ and
$L_I$ are respectively real and imaginary if and only if
    \begin{align*}
    &[L_R,\cT]=0, && \{L_I,\cT\}=0.
    \end{align*}

We can easily show that the real multiples, sums,  and products of
real operators are real. This for instance implies that $\hat
p^2=-(i\hat p)^2$ is a real operator. In light of this observation,
the time-reversal of a standard Hamiltonian operator
(\ref{am-standard-t}) is given by $\overline H(t)=\frac{\hat
p^2}{2m}+\overline{v(\hat x,t)}$, where
    \[\overline{v(\hat x,t)}\phi(x)=
    \cT v(\hat x,-t)\cT\phi(x)=\cT[v(x,-t)\phi(x)^*]=v(x,-t)^*\phi(x).\]
This shows that $v(\hat x,t)$ is a real operator provided that $v(x,-t)=v(x,t)^*$. In particular, for a time-independent standard Hamiltonian,
    \be
    H=\frac{\hat p^2}{2m}+v(\hat x),
    \label{am-standard}
    \ee
we have
    \be
    \overline H=\frac{\hat p^2}{2m}+v(\hat x)^*,
    \label{am-standard-TR}
    \ee
where $v(\hat x)^*\phi(x):=\overline{v(x)}\phi(x)=v(x)^*\phi(x)$.
The  Hamiltonian~(\ref{am-standard}) is therefore real if and only
if $v(x)$ is a real-valued potential.

Next, we explore the consequences of the combined action of
parity and time-reversal transformations. This is realized in
$L^2(\R)$ by $\cP\cT$ whose effect on the wave functions $\phi(x)$
and time-dependent linear operators $L(t)$ are give by
    \bea
    \phi(x)&\longrightarrow&\widetilde{\overline \phi}(x)
    :=(\cP\cT\phi)(x)=\phi(-x,t)^*,\nn\\
    L(t)&\longrightarrow&\widetilde{\overline L}(t):=
    \cP\left[\cT L(-t)\cT^{-1}\right]\cP^{-1}
    =\cP\cT L(-t)(\cP\cT)^{-1}=\cP\cT L(-t)\cP\cT.\nn
    \eea
Here, in the last equality we have used the fact that $\cP$  and
$\cT$ commute and square to identity;
    \begin{align}
    &[\cP,\cT]=0, &&\cP^2=\cT^2=I.
    \label{am-PT-commute}
    \end{align}
Because $\cP$ and $\cT$ are respectively unitary and antiunitary operators,
    \bea
    &&\cP\cT(\alpha_1\phi_1+\alpha_1\phi_2)=
    \cP(\alpha^*\cT \phi_1+\alpha_2^*\cT\phi_2)
    =\alpha^*\cP\cT \phi_1+\alpha_2^*\cP\cT\phi_2,\nn\\
    &&\br\cP\cT\phi_1|\cP\cT\phi_2\kt=\br\cT\phi_1|\cT\phi_2\kt=
    \br\phi_2| \phi_1\kt.\nn
    \eea
These show that $\cP\cT$ is an antiunitary operator. The same
is true about $\cP_a\cT$.

We can use (\ref{am-PT-commute}) and
    \begin{align}
    & \overline{\hat x}=\cT\,\hat x\,\cT^{-1}=\hat x,
    && \overline{\hat p}=\cT\,\hat p\,\cT^{-1}=-\hat p,
    && \widetilde{\hat x}=\cP\,\hat x\,\cP^{-1}=-\hat x,
    && \widetilde{\hat p}=\cP\,\hat p\,\cP^{-1}=-\hat p,
    \label{am-PT-action}
    \end{align}
to show that
    \begin{align}
    & \widetilde{\overline{\hat x}}=\cP\cT\,\hat x\,(\cP\cT)^{-1}=-\hat x,
    &&\widetilde{\overline{\hat p}}=\cP\cT\,\hat p\,(\cP\cT)^{-1}=\hat p.
    \label{am-PT-x-p}
    \end{align}
In other words,
    \begin{align}
    & \{\hat x,\cP\cT\}=0,
    &&[\,\hat p,\cP\cT]=0.
    \end{align}
Another consequence of (\ref{am-PT-x-p}) and the antilinearity  of
$\cP\cT$ is that it transforms a standard Hamiltonian operator of
the form (\ref{am-standard}) to
    \be
    \widetilde{\overline{H}}=\frac{\hat p^2}{2m}+v(-\hat x)^*.
    \label{am-PT-H-trans}
    \ee

A linear operator $L(t)$ is said to be {\em $\cP\cT$-symmetric}  if
it is invariant under the combined action of $\cP$ and $\cT$, i.e.,
$L(t)\to \widetilde{\overline L}(t)=L(t)$. For a time-independent
operator $L$, this means
    \[ [L,\cP\cT]=0.\]
In particular, $\hat p$ is $\cP\cT$-symmetric, and a
time-independent  standard Hamiltonian $H$ is $\cP\cT$-symmetric if
and only if its potential is $\cP\cT$-symmetric, i.e.,
$v(-x)^*=v(x)$. In terms of the real and imaginary parts of $v(x)$,
which we denote by $v_r(x)$ and $v_i(x)$, this condition takes the
form
    \begin{align}
    &v_r(-x)=v_r(x), && v_i(-x)=-v_i(x).
    \end{align}
Therefore, the real and imaginary parts of a $\cP\cT$-symmetric
potential are respectively even and odd functions. Similarly, it
follows that $H$ is $\cP_a\cT$-symmetric if and only if
$v(2a-x)^*=v(x)$. This is equivalent to
    \begin{align}
    &v_r(2a-x)=v_r(x), && v_i(2a-x)=-v_i(x).
    \end{align}

\section{$\cP$-, $\cT$-, and $\cP\cT$-transformation of the scattering data}
\label{am-S7}

Consider the scattering problem for a wave equation in one dimension
that admits solutions $\psi(x)$ satisfying the asymptotic boundary
conditions (\ref{am-pw}). Suppose that for $x\to\pm\infty$ the
parity, time-reversal, and space translations respectively transform
$\psi(x)$ according to:
    \begin{align}
    &\psi(x)\stackrel{\cP}{\longrightarrow}\widetilde\psi(x):=\psi(-x),
    &&\psi(x)\stackrel{\cT}{\longrightarrow}\overline\psi(x):=\psi(x)^*,
    &&\psi(x)\stackrel{T_a}{\longrightarrow}\psi_a(x):=\psi(x-a).
    \label{am-psi-trans}
    \end{align}
It is easy to see that these transformations leave the asymptotic
boundary conditions (\ref{am-pw}) form-invariant. This shows that
the transformed wave functions, $\widetilde\psi(x)$,
$\overline\psi(x)$ and $\psi_a(x)$ also define consistent scattering
problems. We wish to explore the behaviour of the corresponding
reflection and transmission amplitudes. To do this, we confine our 
attention to situations where we can define a transfer matrix $\bM(k)$
and examine the effect of the transformations~(\ref{am-psi-trans}) on 
$\bM(k)$. 

Let $\widetilde\bM(k)$, $\overline\bM(k)$,  and $\bM_a(k)$
respectively  denote the transfer matrix for $\widetilde\psi(x)$,
$\overline\psi(x)$ and $\psi_a(x)$. We can use (\ref{am-pw}),
(\ref{am-M-def}), and (\ref{am-psi-trans}) to related them to
$\bM(k)$. This requires expressing the asymptotic expression for
$\widetilde\psi(x)$, $\overline\psi(x)$ and $\psi_a(x)$  in the form
(\ref{am-pw}) with $(A_\pm,B_\pm)$ respectively replaced by
$(\widetilde A_\pm,\widetilde B_\pm)$, $(\overline A_\pm,\overline
B_\pm)$, and $(A_{a\pm},B_{a\pm})$. In this way we find asymptotic
formulas for $\widetilde\psi(x)$, $\overline\psi(x)$ and $\psi_a(x)$
that together with (\ref{am-psi-trans}) imply:
    \begin{align}
    &\widetilde A_\pm=B_\mp, && \widetilde B_\pm=A_\mp,
    \label{am-asym-P}\\
    &\overline A_\pm=B_\pm^*, && \overline B_\pm=A_\pm^*,
    \label{am-asym-T}\\
    &A_{a\pm}=e^{-iak}A_\pm, && B_{a\pm}=e^{iak} B_\pm.
    \label{am-asym-Tb}
    \end{align}
Recalling that the transfer matrices $\widetilde\bM$,
$\overline\bM$, and $\bM_a$ satisfy
    \begin{align}
    &\left[\begin{array}{c}
    \widetilde A_+\\
    \widetilde B_+\end{array}\right]=\widetilde\bM\left[\begin{array}{c}
    \widetilde A_-\\
    \widetilde B_-\end{array}\right],
    &&\left[\begin{array}{c}
    \overline A_+\\
    \overline B_+\end{array}\right]=\overline\bM\left[\begin{array}{c}
    \overline A_-\\
    \overline B_-\end{array}\right],
    &&\left[\begin{array}{c}
    A_{a+}\\
    B_{a+}\end{array}\right]=\bM_a\left[\begin{array}{c}
     A_{a-}\\
     B_{a-}\end{array}\right],
    \label{am-transfer-matrices}
    \end{align}
we can use (\ref{am-M-def}) and (\ref{am-asym-P}) -- (\ref{am-asym-Tb}) to infer:
    \begin{align}
    & \widetilde\bM=\bsigma_1\bM^{-1}\bsigma_1,
    &&\overline\bM=\bsigma_1\bM^*\bsigma_1,
    && \bM_a=e^{-iak\bsigma_3}\bM\, e^{iak\bsigma_3},
    \label{am-trans-M=}
    \end{align}
where
    \begin{align*}
    &\bsigma_1:=\left[\begin{array}{cc}
    0 & 1\\
    1 & 0\end{array}\right],
    %&&\bsigma_2:=\left[\begin{array}{cc}
    %0 & -i\\
    %i & 0\end{array}\right],
    &&\bsigma_3:=\left[\begin{array}{cc}
    1 & 0\\
    0 & -1\end{array}\right],
    &&e^{ia\bsigma_3}=\left[\begin{array}{cc}
    e^{ia} & 0\\
    0 & e^{-ia}\end{array}\right].
    \end{align*}

It is instructive to examine the explicit expression for the
entries of $\widetilde\bM$, $\overline\bM$,  and $\bM_a$ . According
to (\ref{am-trans-M=}), they have the form:
    \begin{align}
    &\widetilde M_{11}=\frac{M_{11}}{\det\bM},
    &&\widetilde M_{12}=-\frac{M_{21}}{\det\bM},
    &&\widetilde M_{21}=-\frac{M_{12}}{\det\bM},
    &&\widetilde M_{22}=\frac{M_{22}}{\det\bM},
    \label{am-widetilde-M}\\
    &\overline M_{11}=M_{22}^*,
    &&\overline M_{12}=M_{21}^*,
    &&\overline M_{21}=M_{12}^*,
    &&\overline M_{22}=M_{11}^*,
    \label{am-overline-M}\\
    & M_{a11}=M_{11},
    && M_{a12}=e^{-2iak}M_{12},
    && M_{a21}=e^{2iak}M_{21},
    && M_{a22}=M_{22}.
    \label{am-Mb}
    \end{align}
We can use these relations together with (\ref{am-rt=M}) to  compute
the reflection and transmission amplitudes for the reflected,
time-reversed, and translated waves, $\widetilde\psi(x)$,
$\overline\psi(x)$, and $\psi_a(x)$. These are respectively given by
    \begin{align}
    & \widetilde \fr_l= \fr_r ,
    &&\widetilde \ft_l=\ft_r,
    &&\widetilde \fr_r=\fr_l,
    &&\widetilde \ft_r=\ft_l,
    \label{am-P-trans-ampl}\\
    & \overline \fr_l=-\frac{\fr_r^*}{\fD^*},
    &&\overline \ft_l=\frac{\ft_l^*}{\fD^*},
    &&\overline \fr_r=-\frac{\fr_l^*}{\fD^*},
    &&\overline \ft_r=\frac{\ft_r^*}{\fD^*},
    \label{am-T-trans-ampl}\\
    & \fr_{al}=e^{2iak}\,\fr_l,
    && \ft_{al}=\ft_l
    && \fr_{ar}=e^{-2iak}\,\fr_r,
    && \ft_{ar}=\ft_r,
    \label{am-Ta-trans-ampl}
    \end{align}
where we recall that $\fD:= M_{11}/M_{22}=\ft_l\ft_r-\fr_l\fr_r=\det\bS$.

Next, we examine the effect of $\cP_a$ on the scattering data.
Because in view of (\ref{am-parity=}) we have $\cP_a=T_{2a}\cP$,
$\cP_a$ transforms the transfer matrix $\bM$ according to
    \bea
    \bM\stackrel{\cP_a}{\longrightarrow} \widetilde{M}_{\,2a}&=&
     e^{-i2ak\bsigma_3}\bsigma_1 \bM^{-1}\bsigma_1 e^{i2ak\bsigma_3}
     %\nn\\&=&
     =\frac{1}{\det\bM}\left[\begin{array}{cc}
    M_{11} & -e^{-4iak}M_{21}\\
    -e^{4aik}M_{12} & M_{22}\end{array}\right].
    \label{am-Pa-M-transform}
    \eea
Here we have made use of (\ref{am-trans-M=}) and the identity
    \[e^{-i\varphi\bsigma_3}\bsigma_1=\bsigma_1 e^{i\varphi\bsigma_3}=
    \left[\begin{array}{cc}
    0 & e^{-i\varphi}\\
    e^{i\varphi}& 0\end{array}\right].\]
Using  (\ref{am-rt=M}) and (\ref{am-Pa-M-transform}), we obtain
    \begin{align}
    &\fr_l\stackrel{\cP_a}{\longrightarrow} e^{4iak}\fr_r=e^{4iak}\widetilde\fr_l,
    &&\ft_l\stackrel{\cP_a}{\longrightarrow} \ft_r= \widetilde\ft_r,
    \label{am-Pa-trans-left}\\
    &\fr_r\stackrel{\cP_a}{\longrightarrow} e^{-4iak}\fr_l=e^{-4iak}\widetilde\fr_r,
    &&\ft_r\stackrel{\cP_a}{\longrightarrow} \ft_l=\widetilde\ft_r.
    \label{am-Pa-trans-right}
    \end{align}
These equations show that the effect of a space-reflection  about a
point $a\neq 0$ introduces the extra phase factors $e^{\pm4iak}$ in
the expression for the $\cP$-transformed reflection amplitudes. In
particular, it does not affect the zeros and singularities of the
reflection and transmission amplitudes of the system.

We now study the implication of $\cP\cT$ on the scattering  data.
According to (\ref{am-trans-M=}) the $\cP\cT$-transformation of the
transfer matrix $\bM(t)$ yields
    \bea
    \bM\stackrel{\cP\cT}{\longrightarrow}
    \widetilde{\overline\bM}&=&\bsigma_1[\bsigma_1\bM^*\bsigma_1]^{-1}
    \bsigma_1=\bM^{-1*}\nn\\
    &=&\frac{1}{\det\bM^*}
    \left[\begin{array}{cc}
    M_{22}^* & -M_{12}^*\\
    -M_{21}^* & M_{11}^*\end{array}\right].
    \label{am-PT-M-transform}
    \eea
In particular,
    \be
    \det\bM\stackrel{\cP\cT}{\longrightarrow}\det\widetilde{\overline\bM}
    =\frac{1}{\det\bM^*},
    \label{am-PT-detM-transform}
    \ee
    \begin{align}
    & M_{11}\stackrel{\cP\cT}{\longrightarrow}\widetilde{\overline{M}}_{11}:=
    \frac{M_{22}^*}{\det\bM^*},
    && M_{12}\stackrel{\cP\cT}{\longrightarrow}\widetilde{\overline{M}}_{12}:=
    -\frac{M_{12}^*}{\det\bM^*},
    \label{am-PT-M1-transform}\\
    & M_{21}\stackrel{\cP\cT}{\longrightarrow}\widetilde{\overline{M}}_{21}:=
    -\frac{M_{21}^*}{\det\bM^*},
    && M_{22} \stackrel{\cP\cT}{\longrightarrow} \widetilde{\overline{M}}_{22}:=
    \frac{M_{11}^*}{\det\bM^*}.
    \label{am-PT-M2-transform}
    \end{align}
With the help of these relations and (\ref{am-rt=M}) or
alternatively  (\ref{am-P-trans-ampl}) and (\ref{am-T-trans-ampl}),
we can derive the following expressions for the $\cP\cT$-transformed
reflection and transmission amplitudes.
    \begin{align}
    &\widetilde{\overline{\fr}}_{l}=-\frac{\fr_{l}^*}{\fD^*},
    &&\widetilde{\overline{\ft}}_{l}=\frac{\ft_{r}^*}{\fD^*},
    &&\widetilde{\overline{\fr}}_{r}=-\frac{\fr_{r}^*}{\fD^*},
    &&\widetilde{\overline{\ft}}_{r}=\frac{\ft_{l}^*}{\fD^*}.
    \label{am-PT-rt-transform}
    \end{align}

\section{$\cP$-, $\cT$-, and $\cP\cT$-symmetric scattering systems}
\label{am-Sec8}

A physical system that involves the scattering of a scalar wave in
one  dimension is said to be {\em $\cP$-,  $\cT$-, and
$\cP\cT$-symmetric} if its reflection and transmission amplitudes
are respectively invariant under space-reflection, time-reversal,
and the combined action of space-reflection and time-reversal
transformation, i.e.,
    \begin{align}
    \mbox{$\cP$-symmetry}&:=~ \widetilde\fr_{l/r}=\fr_{l/r}~~{\rm and}~~\widetilde\ft_{l/r}=
    \ft_{l/r},
    \label{am-P-sym-def}\\
    \mbox{$\cT$-symmetry}&:=~ \overline\fr_{l/r}=\fr_{l/r}~~{\rm and}~~\overline\ft_{l/r}=
    \ft_{l/r},
    \label{am-T-sym-def}\\
    \mbox{$\cP\cT$-symmetry}&:=~\widetilde{\overline\fr}_{l/r}=\fr_{l/r}~~{\rm and}~~\          \widetilde{\overline\ft}_{l/r}=\ft_{l/r}.
    \label{am-PT-sym-def}
    \end{align}
We can alternatively state the definition of these symmetries in
terms of  the invariance of the transfer matrix $\bM$ or the
scattering matrix $\bS$ of the system under the action of $\cP$,
$\cT$, and $\cP\cT$. In this section we explore the consequences of
these symmetries.

According to (\ref{am-P-trans-ampl}), the $\cP$-symmetry of a
scattering  system implies         \begin{align}
    &\fr_l=\fr_r, && \ft_l=\ft_r.
    \label{am-P-sym-r}
    \end{align}
Substituting the latter equation in (\ref{am-det-M=}), we find
$\det\bM=1$.  Let us also mention that in view of
(\ref{am-S-eigenvalues-gen}) and (\ref{am-P-sym-r}), the eigenvalues
of the $\bS$-matrix for $\cP$-symmetric systems take the simple
form: $\fs_\pm=\ft\pm\fr$ where $\ft:=\ft_l=\ft_r$ and
$\fr:=\fr_l=\fr_r$. Another obvious consequence of
(\ref{am-P-sym-r}) is that $\cP$-symmetric systems cannot support
unidirectional reflection or unidirectional invisibility.

The delta-function potential~(\ref{am-delta-potential}) provides a
simple  example of a $\cP$-symmetric potential that may not be
time-reversal-invariant. As demonstrated by (\ref{am-r-t-delta}), it
complies with (\ref{am-P-sym-r}).

We can similarly derive the consequences of $\cP_a$-symmetry. This
symmetry also implies transmission reciprocity and $\det\bM=1$, but
breaks the reciprocity in reflection amplitudes as it yields the
following generalization of the first relation in
(\ref{am-P-sym-r}).
    \be
    e^{-2iak}\fr_l(k)=e^{2aik}\fr_r(k).
    \label{am-Pa-sym-ref}
    \ee
Notice however that reciprocity in reflection coefficients,
$|\fr_l|^2=|\fr_r|^2$, persists. A simple example of
$\cP_a$-symmetric scattering system is that of the barrier potential
(\ref{am-barrier}) with $L=2a$. Clearly in this case the expressions
(\ref{am-barrier-r-left}) and (\ref{am-barrier-r-right}) for the
reflection amplitudes agree with (\ref{am-Pa-sym-ref}).

The consequences of the $\cT$-symmetry are more interesting.
Imposing (\ref{am-T-sym-def}), we can use (\ref{am-T-trans-ampl}) to
deduce
    \begin{align}
    &\fr_r^*=-\fD^*\fr_l, && \fr_l^*=-\fD^*\fr_r, && \ft_{l/r}^*=\fD^*\ft_{l/r}.
    \label{am-T-sym-rt}
    \end{align}
The first two of these relations indicate that either both
$\fr_{l/r}$  vanish or $|\fD|=1$. This means that there is some real
number $\sigma\in\R$ such that $\fD=e^{i\sigma}$. Substituting this
in (\ref{am-T-sym-rt}), we can show that
    \begin{align}
    &\fr_r=-e^{i\sigma}\fr_l^*, && \ft_{l/r}=\epsilon_{l/r}|\ft_{l/r}|e^{i\sigma/2},
    \label{am-T-sym-rt-2}
    \end{align}
where $\epsilon_{l/r}$ are some unspecified signs;
$\epsilon_{l/r}\in\{-1,1\}$.  In particular,
    \be
    |\fr_l|=|\fr_r|.
    \label{am-ref-reciprocity}
    \ee
This equation proves the following result.
    \begin{theorem}
    Time-reversal-invariant systems in one dimension cannot support
    unidirectional reflection or unidirectional invisibility.
    \label{am-thm-no-unidir}
    \end{theorem}

If we insert (\ref{am-T-sym-rt-2}) in the definition of $\fD$,
namely  (\ref{am-fD-def}), and impose $\fD=e^{i\sigma}$, we find
    \be
    |\fr_l|^2+\epsilon_l\epsilon_r|\ft_l\ft_r|=1.
    \label{am-T-unitarity}
    \ee
The following theorem summarizes the content of
Equations~(\ref{am-ref-reciprocity}) and (\ref{am-T-unitarity}).
    \begin{theorem}
    The reflection and transmission amplitudes of a
    time-reversal-invariant scattering system
    in one-dimension satisfy
        \be
        |\fr_l(k)|^2=|\fr_r(k)|^2=1\pm|\ft_l(k)\ft_r(k)|,
        \label{am-T-sym-thm}
        \ee
    where $k\in\R^+$ and the unspecified sign on the right-hand side
    is to be taken negative whenever the system has reciprocal transmission,
    i.e., $\ft_l(k)=\ft_r(k)$.
    \label{am-thm7}
    \end{theorem}

If $\ft_r=\ft_l$, which is for example the case for systems that are
both $\cT$- and $\cP$-symmetric or described by a real scattering
potential,  $\epsilon_l=\epsilon_r$ and we can write
(\ref{am-T-unitarity}) as
    \be
    |\fr|^2+|\ft|^2=1,
    \label{am-unitarity}
    \ee
where again $\ft:=\ft_l=\ft_r$ and $\fr:=\fr_l=\fr_r$.
Equation~(\ref{am-unitarity}) is usually derived for real scattering
potentials using the unitarity of the time-evolution generated by
the corresponding standard Hamiltonian (\ref{am-standard}). It is
therefore often called the {\em unitarity relation}. The derivation
we have offered here is more general, for it relies on the
transmission reciprocity and time-reversal-invariance. Removing the
first of these conditions, we arrive at (\ref{am-T-sym-thm}) which
is a mild generalization of the unitarity relation
(\ref{am-unitarity}). Equations (\ref{am-T-sym-thm}) apply, for
example, to the scattering problem defined by the time-independent
Schr\"odinger equation for the Hamiltonian operator:
    \[ H= (I+e^{-\mu\hat x^2})\left[ \frac{\hat p^2}{2m} +v(\hat x)\right],\]
where $\mu$ is a positive real parameter, and $v(x)$ is a  real and
even scattering potential. Note that this Hamiltonian is both $\cP$-
and $\cT$-symmetric but not Hermitian.\footnote{The scattering
problem for this Hamiltonian operator is equivalent to that of the
energy-dependent scattering potential $v(x,k):=2m v(x)+
k^2/(1+e^{\mu x^2})$. This is because we can write $H\psi(x)=E\psi(x)$
in the form  $-\psi''(x)+v(x,k)\psi(x)=k^2\psi(x)$ where $k:=\sqrt E$.}

The unitarity relation~(\ref{am-unitarity}), which holds for
time-reversal-invariant systems with reciprocal transmission, in
general, and real scattering potentials in particular, implies that
the reflection and transmission coefficients of the system cannot
exceed 1; $|\fr(k)|^2\leq 1$ and $|\ft(k)|^2\leq 1$ for all
$k\in\R^+$. This means that {\em these systems do not amplify the
transmitted or reflected waves}. In particular, we have:
    \begin{theorem}
    If a time-reversal-invariant scattering system in one dimension has
    reciprocal transmission, it
    cannot have spectral singularities.
    \label{am-thm-no-SS}
    \end{theorem}
It is for this reason that spectral singularities do not appear  in
the study of unitary quantum systems described by standard
Hamiltonian operators.

Time-reversal-invariant systems violating reciprocity in
transmission can have spectral singularities. A simple example is a
single-center point interaction (\ref{am-pt-int}) with $n=1$,
$c_1=0$, and
    \be
    \bB=\left[\begin{array}{cc}
    \alpha & \beta\\
    \gamma & -\alpha\end{array}\right],\quad\quad\alpha,\beta,\gamma\in\R,~~\beta\gamma>0.
    \ee
It is easy to see that the system described by this point
interaction is time-reversal-invariant, because $\bB$ is a real
matrix \cite{am-jpa-2011}. Furthermore, we can compute its transfer
matrix using (\ref{am-pt-int-M=}) and find out that for this system
$M_{22}(k)=\beta k^2-\gamma$. Therefore, it has a spectral singular
$k_0^2=\gamma/\beta$.  Note also that because
$\det\bM=\det\bB=-\alpha^2-\beta\gamma<0$, $\det\bM\neq 1$ which
shows that it has nonreciprocal transmission.

We can also characterize time-reversal symmetry in terms of the restrictions it imposes on the transfer and scattering matrices. These have the following simple form.
    \begin{align}
    &\bM^*=\bsigma_1\bM\bsigma_1,
    &&\bS^*=\bsigma_1\bS^{-1}\bsigma_1.
    \label{am-T-sym-M-S}
    \end{align}
Because $\det\bsigma_1=-1$, the first of these equations implies that $\det\bM$ must be real while
the second reproduces the result that $\det\bS$ is unimodular; $|\det\bS|=1$.

Let us examine the eigenvalues of $\bS$ for time-reversal-invariant systems. In view of (\ref{am-S-eigenvalues-gen}), (\ref{am-T-sym-rt-2}), and (\ref{am-T-unitarity}), these are given by
    \begin{align}
    &\fs_+=(\tau+\sqrt{\tau^2-1})e^{i\sigma/2},
    &&\fs_-=(\tau-\sqrt{\tau^2-1})e^{i\sigma/2}=
    \frac{e^{i\sigma/2}}{\tau+\sqrt{\tau^2-1}},
    \label{am-spm-T-sym}
    \end{align}
where
    \be
    \tau(k):=\frac{\epsilon_l|\ft_l(k)|+\epsilon_r|\ft_r(k)|}{2}.
    \label{am-tau=}
    \ee
It is not difficult to see that $|\fs_\pm|=1$ if and only if
    \be
    |\tau|\leq 1.
    \label{am-T-sym-exact-0}
    \ee
If $|\tau(k)|\leq 1$ for all $k\in\R^+$, we say that the time-reversal symmetry of the system is {\em exact} or {\em unbroken}. If $|\tau|>1$ for some $k\in\R^+$, we say that the system has a {\em broken time-reversal symmetry}.

To examine the physical meaning of exact time-reversal symmetry, we examine the consequences of (\ref{am-T-sym-exact-0}). First we use (\ref{am-tau=}) to write it in the form
    \be
    |\ft_l|^2+|\ft_r|^2+2\epsilon_l\epsilon_r|\ft_l\ft_r|\leq 4.
    \label{am-T-sym-exact-1}
    \ee
With the help of (\ref{am-T-unitarity}), we can express this equation as
    \be
    \frac{|\ft_l|^2+|\ft_r|^2}{2}\leq  1+|\fr_l|^2 .
    \label{am-T-sym-exact-2}
    \ee
If $\epsilon_l\epsilon_r=1$, (\ref{am-T-unitarity}), (\ref{am-ref-reciprocity}), and (\ref{am-T-sym-exact-2}) imply
    \begin{align}
    &|\fr_{l/r}|^2\leq 1,  && |\ft_l|^2+|\ft_r|^2\leq 4.
    \end{align}
Therefore similarly to the unitary systems the reflection and transmission amplitudes are bounded functions, and the system cannot involve spectral singularities.

If a system has a broken time-reversal symmetry, there is some $k\in\R^+$ such that
$|\tau(k)|>1$. In this case, (\ref{am-T-unitarity}) implies
    \be
    \frac{|\ft_l(k)|^2+|\ft_r(k)|^2}{2}|>1+|\fr_l|^2\geq 1.
    \label{am-T-sym-broken-1}
    \ee
Furthermore because $\sqrt{\tau^2-1}$ is real and nonzero, (\ref{am-spm-T-sym}) implies $|\fs_\pm|\neq 1$ and $\fs_-=1/\fs_+^*$.

Equation~(\ref{am-T-unitarity}) that reveals various properties of the time-reversal-invariant scattering systems has a rather interesting equivalent that does not involve the unspecified signs $\epsilon_{l/r}$. To derive this, first we use (\ref{am-btrvr-rt}) and the fact that $\fD(k)=e^{i\sigma(k)}$ to show that
    \begin{align}
    &\fr_{l/r}(-k)=-e^{-i\sigma(k)}\fr_{r/l}(k),
    &&\ft_{l/r}(-k)=e^{-i\sigma(k)}\ft_{l/r}(k).
    \label{am-T-sym-minus-k}
    \end{align}
These relations have the following straightforward implications:
    \be
    |\fr_{l/r}(-k)|=|\fr_{r/l}(k)|,\quad\quad\quad\quad |\ft_{l/r}(-k)|=|\ft_{l/r}(k)|,
    \label{am-T-sym-rel-x1}
    \ee
    \be
    \fr_{l/r}(-k)\fr_{l/r}+\ft_{l/r}(-k)\ft_{r/l}(k)=1,
    \label{am-T-sym-rel-x2}
    \ee
where we have made use of the definition of $\fD(k)$, i.e., (\ref{am-fD-def}), and the fact that $\fD(k)=e^{i\sigma(k)}$.

It is important to notice that our derivation of Equations (\ref{am-T-sym-minus-k}) -- (\ref{am-T-sym-rel-x2}) only uses the fact that $|\fD(k)|=1$, which is much less restrictive than the time-reversal symmetry of the system. We state this result as a theorem:
    \begin{theorem}
    Equations (\ref{am-T-sym-rel-x1}) and (\ref{am-T-sym-rel-x2}) hold for
    any scattering system whose reflection and transmission amplitudes satisfy
    $|\ft_l(k)\ft_r(k)-\fr_l(k)\fr_r(k)|=1$, i.e., $|\fD(k)|=1$.\footnote{An extension of this theorem to more general scattering systems is given in \cite{p145}.}
    \label{am-thm-rel-x}
    \end{theorem}

Next, we examine the implications of $\cP\cT$-symmetry. In view of  (\ref{am-PT-rt-transform}) and (\ref{am-PT-sym-def}), the reflection and transmission amplitudes of $\cP\cT$-symmetric scattering systems satisfy
    \begin{align}
    &\fr_{l/r}^*=-\fD^*\fr_{l/r}, && \ft_{l/r}^*=\fD^*\ft_{l/r}.
    \label{am-PT-sym-rt}
    \end{align}
If we complex-conjugate both sides of (\ref{am-fD-def}) and     use (\ref{am-PT-sym-rt}) in the right-hand side of the resulting equation, we find $\fD^*=\fD^{*2}\fD$, which means $|\fD|=1$. In view of Theorem~\ref{am-thm-rel-x}, this shows that that, similarly to time-reversal-invariant systems, $\cP\cT$-symmetric systems satisfy the identities (\ref{am-T-sym-rel-x1}) and (\ref{am-T-sym-rel-x2}).\footnote{Equations~(\ref{am-T-sym-rel-x1}) was originally conjectures in \cite{am-ahmed} for $\cP\cT$-symmetric scattering potentials based on evidence provided by the study of a complexified Scarf II potential. It was subsequently proven in \cite{am-jpa-2014c} for general $\cP\cT$-symmetric scattering potentials which respect transmission reciprocity.}

Because $|\fD|=1$, $\fD=e^{i\sigma}$ for some $\sigma\in\R$. Using this relation in  (\ref{am-PT-sym-rt}), we can show that
    \begin{align}
    &\fr_{l/r}=i\eta_{l/r}  e^{i\sigma/2}|\fr_{l/r}|,
    &&\ft_{l/r}=\epsilon_{l/r} e^{i\sigma/2}|\ft_{l/r}|,
    \label{am-PT-sym-rt=}
    \end{align}
where $\eta_{l/r},\epsilon_{l/r}\in\{-1,1\}$. Now, we substitute these relations in (\ref{am-fD-def}) and make use of $\fD=e^{i\sigma}$ to conclude that
    \be
    \epsilon_l\epsilon_r|\ft_l\ft_r|+\eta_l\eta_r|\fr_l\fr_r|=1.
    \label{am-PT-unitarity}
    \ee
According to this equation, $\epsilon_l\epsilon_r=-1$ implies $\eta_l\eta_r=1$ and $\eta_l\eta_r=-1$ implies $\epsilon_l\epsilon_r=1$. These observations prove the following theorem.
    \begin{theorem}
    For all $k\in\R^+$, the reflection and transmission amplitudes of a $\cP\cT$-symmetric scattering system in one-dimension satisfy either
        \bea
        &&|\ft_l(k)\ft_r(k)|=-1+|\fr_l(k)\fr_r(k)|,
        \label{am-PT-sym-thm8-1}
        \eea
    or
        \bea
        &&|\ft_l(k)\ft_r(k)|=1\pm|\fr_l(k)\fr_r(k)|.
        \label{am-PT-sym-thm8-2}
        \eea
    If the system has reciprocal transmission, i.e., $\ft_l(k)=\ft_r(k)$, only the second of these relations holds. In this case, we have
        \be
        |\ft(k)|^2\pm |\fr_l(k)\fr_r(k)|=1.
        \label{am-PT-sym-thm8-3}
        \ee
    If the system has reciprocal reflection, i.e., $\fr_l(k)=\fr_r(k)$, (\ref{am-PT-sym-thm8-1}) is not excluded but the unspecified sign on the right-hand side of (\ref{am-PT-sym-thm8-2}) is to be   taken negative, i.e., it reads
        \be
        |\ft_l(k)\ft_r(k)|+ |\fr(k)|^2=1.
        \label{am-PT-sym-thm8-4}
        \ee
    \label{am-thm8}
    \end{theorem}
For a scattering system defined by a $\cP\cT$-symmetric scattering potential, Theorem~\ref{am-thm1} ensures the reciprocity in transmission. Therefore, $\cP\cT$-symmetric scattering potentials satisfy (\ref{am-PT-sym-thm8-3}), \cite{stone-2012}.

Next, we examine the effect of $\cP\cT$-symmetry on the transfer and scattering matrices. It is easy to show that for $\cP\cT$-symmetric systems,
    \begin{align}
    &\bM^*=\bM^{-1},
    &&\bS^\dagger=\bsigma_1\bS^{-1}\bsigma_1,
    \label{am-PT-sym-psudo-unitary}
    \end{align}
where $\bS^\dagger$ is the conjugate-transpose or Hermitian-conjugate of $\bS$. The first of these relations follows from (\ref{am-PT-M-transform}) and implies that $\det\bM$ is unimodular;
    \be
    |\det\bM|=1.
    \ee
The second is a consequence of (\ref{am-S-matrix=}) and (\ref{am-PT-sym-rt}). Because $\bsigma_1^{-1}=\bsigma_1$, we can write it in the form $\bS^\dagger=\bsigma_1\bS^{-1}\bsigma_1^{-1}$. This indicates that $\bS$ is a {\em $\bsigma_1$-pseudo-unitary matrix} \cite{am-jmp-2004}, i.e., if we identify the elements of
$\C^2$ with $2\times 1$ matrices and view $\bsigma_1$ and $\bS$ as linear operators acting on them, then $\bS$ preserves the indefinite inner product:
    \[ \br \bba,\bb\kt_{\bsigma_1}:=\br\bba|\bsigma_1\bb\kt=\bba^\dagger\bsigma_1\bb=
    a_1^*b_2+a_2^*b_1,\]
where $\bba=[a_1~a_2]^T$ and $\bb=[b_1~b_2]^T$ are arbitrary $2\times 1$ complex matrices, and a superscript ``T'' on a matrix labels its transpose.\footnote{For a $2\times2$ matrix $\bA$, the condition of being $\bsigma_1$-pseudo-unitary is equivalent to the requirement that $e^{i\pi\bsigma_2/4}\bA e^{-i\pi\bsigma_2/4}$ belong to the pseudo-unitary group $U(1,1)$, where $\bsigma_2$ is the second Pauli matrix.} Because the $\bS$-matrix of $\cP\cT$-symmetric scattering potentials are $\bsigma_1$-pseudo-unitary, Equation~(\ref{am-PT-sym-thm8-3}) is sometimes called the {\em pseudo-unitarity relation}.

In general, an invertible square matrix $\bcU$ is said to be {\em pseudo-unitary}, if there is an invertible Hermitian matrix $\bfeta$ such that $\bcU^\dagger=\bfeta\,\bcU^{-1}\bfeta^{-1}$. Pseudo-unitary matrices have the property that the inverse of the complex-conjugate of their eigenvalues are also eigenvalues, i.e., if $\fs$ is an eigenvalue of a pseudo-unitary matrix, either $|\fs|=1$ or $1/\fs^*$ is also an eigenvalue \cite{am-jmp-2004}. As we show above this condition applies also for the eigenvalues of the $\bS$-matrix for time-reversal-invariant systems. We can check its validity for the $\bS$-matrix of $\cP\cT$-symmetric systems by a direct calculation of its eigenvalues. Inserting  (\ref{am-PT-sym-rt=}) in (\ref{am-S-eigenvalues-gen}) and making use of (\ref{am-PT-unitarity}), we find that the expression for $\fs_\pm$ coincides with the one we obtain for the time-reversal-invariant systems, namely (\ref{am-spm-T-sym}). Therefore, again either $|\tau|\leq 1$ in which case $|\fs_\pm|=1$, or $|\tau|>1$ in which case $|\fs_\pm|\neq 1$ and $\fs_-=1/\fs_+^*$.

Following the terminology we employed in our discussion of time-reversal symmetry, we use
the sign of $1-|\tau|$ to introduce the notions of exact and broken $\cP\cT$-symmetry.  If for all $k\in\R^+$, $1-\tau(k)|\geq 0$ so that $|\fs_\pm(k)|=1$, we say that the system has an {\em exact or unbroken $\cP\cT$-symmetry.} If this is not the case we say that its $\cP\cT$-symmetry is {\em broken}. This terminology should not be confused with the one employed in the study of $\cP\cT$-symmetric Hamiltonian operators $H$ that have a discrete spectrum. For these systems unbroken $\cP\cT$-symmetry means the existence of a complete set of eigenvectors of $H$ that are also eigenvectors of $\cP\cT$. This in turn implies the reality of the spectrum of $H$, \cite{bender-1998}. Scattering theory for a $\cP\cT$-symmetric Hamiltonian is sensible only if its spectrum contains a real continuous part that covers the positive real axis in the complex plane. In particular it may or may not have nonreal eigenvalues.\footnote{We use the term ``eigenvalue'' to mean an element of the point spectrum of $H$ which has a square-integrable eigenfunction.}

If for some $k\in\R^+$, a $\cP\cT$-symmetric system has reciprocal transmission, $\tau(k)=|\ft(k)|$. Therefore the condition $|\tau|\leq 1$ puts an upper bound of $1$ on the transmission coefficient $|\ft(k)|^2$. This in turn implies that the unspecified sign in (\ref{am-PT-sym-thm8-3}) must be taken positive and $|\fr_l(k)\fr_r(k)|\leq 1$. As a result, the system cannot amplify reflected or transmitted waves having wavenumber $k$. In particular $k^2$ cannot be a spectral singularity. In summary, {\em for a system with reciprocal transmission, such as those described by a scattering potential, exactness of $\cP\cT$-symmetry forbids amplification of the reflected and transmitted waves and spectral singularities.}

An important advantage of $\cP\cT$-symmetry over $\cP$- and $\cT$-symmetries, is that it does not imply the equality of the left and right reflection amplitudes. Therefore {\em unidirectional reflection and unidirectional invisibility are not forbidden by $\cP\cT$-symmetry}. In fact, it turns out that it is easier to achieve unidirectional reflectionlessness and invisibility in the presence of $\cP\cT$-symmetry than in its absence. This has to do with the following result that is a straightforward consequence of (\ref{am-PT-rt-transform}).
    \begin{theorem}
    The equations characterizing unidirectional invisibility, namely
    \begin{align}
    &\fr_{l/r}(k)=0\neq\fr_{r/l},
    &&\ft_{l/r}(k)=1,
    \label{am-unidir-gen}
    \end{align}
    are invariant under the $\cP\cT$-transformation.
    \end{theorem}
Notice that the statement of this theorem holds also for systems violating transmission reciprocity.

For a $\cP\cT$-symmetric system the equations of unidirectional invisibility enjoy the same symmetry as that of the underlying wave equation. This leads to enormous practical simplifications in constructing specific unidirectionally invisible models. It does not however imply that $\cP\cT$-symmetry is a necessary condition for unidirectional reflection or invisibility \cite{am-pra-2013}.

\section{Time-reversed and self-dual spectral singularities}
\label{am-Sec9}

Consider a linear scattering system $\cS$ with an invertible transfer matrix $\bM(k)$. Then spectral singularities of this system are determined by the real and positive zeros of $M_{22}(k)$.  According to (\ref{am-overline-M}), $M_{11}(k)=0$ if and only if $\overline M_{22}(k)=0$. This in turn means that the real and positive zeros of $M_{11}(k)$ give the spectral singularities of the time-reversed system $\overline{\cS}$. We will refer to these as the {\em time-reversed spectral singularities} of $\cS$.

At a time-reversed spectral singularity the Jost solutions of the time-reversed system become linearly dependent and satisfy purely outgoing boundary conditions at $x=\pm\infty$. This suggests the presence of solutions of the wave equation for the system $\cS$ that satisfy purely incoming asymptotic boundary conditions. To see this, first we note that according to Equation (\ref{am-M-def}) whenever $M_{11}(k)=0$, we can have a solution $\psi(x)$ of the wave equation satisfying (\ref{am-pw}) with $A_+(k)=B_-(k)=0$, i.e.,
    \be
    \psi(x)\to N_\pm(k) e^{\mp ikx} \quad\quad{\rm for} \quad\quad x\to\pm\infty,
    \label{am-incoming}
    \ee
where $N_\pm(k)$ are nonzero complex coefficients satisfying
    \be
    N_+(k)=M_{21}(k)N_-(k).
    \label{am-amp-match}
    \ee
In other words, $\psi(x)$ satisfies the asymptotic boundary conditions~(\ref{am-pw}) with
        \begin{align*}
        &A_-(k)=N_-(k), && B_-(k)=0, && A_+(k)=0, && B_+(k)=N_+(k).
        \end{align*}
If we substitute these in the first equation in (\ref{am-s-matrices}) and recall that $\bS_1=\bS$, we find that
    \[\bS(k)\left[\begin{array}{c}
    N_-(k)\\ N_+(k)\end{array}\right]=\left[\begin{array}{c}
    0\\ 0\end{array}\right].\]
This shows that $[N_-(k)~N_+(k)]^T$ is an eigenvector of $\bS(k)$ with eigenvalue zero. In particular, one of the eigenvalues of $\bS(k)$ vanishes.

The existence of a solution of the wave equation having the asymptotic expression (\ref{am-incoming}) means that the scatterer will absorb any pair of incident left- and right-going waves whose complex amplitude $N_\pm(k)$ are related by (\ref{am-amp-match}). This phenomenon is called {\em coherent perfect absorption} \cite{am-chong-2010,am-longhi-2010-cpa,am-wan}. In the study of effectively one-dimensional optical systems, spectral singularities correspond to the initiation of laser oscillations in a medium with gain, i.e., a laser, while their time-reversal give rise to perfect absorption of finely tuned coherent incident beams by a medium with loss. The latter is sometimes called an {\em antilaser}.

It may happen that a particular wavenumber $k_0$ is a common zero of both $M_{11}(k)$ and $M_{22}(k)$. In this case, we call $k_0^2$ a {\em self-dual spectral singularity} \cite{am-jpa-2012}. At a self-dual spectral singularity the wave equation admits both purely outgoing and purely incoming solutions. This means that if the system is not subject to any incident wave, it will amplify the background noise and begin emitting outgoing waves of wavenumber $k_0$. But if it is  subject to a pair of left- and right-going incident waves with wavenumber $k_0$ and complex amplitudes satisfying (\ref{am-amp-match}) for $k=k_0$, then it will absorb them completely. In its optical realizations this corresponds to a special laser that becomes a coherent perfect absorber (CPA) once it is subject to an appropriate pair of incoming waves. Such a device is called a {\em CPA-laser}.

For a time-reversal-invariant system we have $M_{11}(k)=M_{22}(k)^*$. Therefore every spectral singularity is self-dual. But according to Theorem~\ref{am-thm-no-SS} spectral singularities are forbidden for time-reversal-invariant systems with reciprocal transmission. This excludes real scattering potentials. There are however nonreal potentials that admit self-dual spectral singularities. Principal examples are $\cP\cT$-symmetric scattering potentials \cite{am-longhi-2010-CPA-laser,am-chong-2011,am-wong-2016}. According to (\ref{am-PT-M1-transform}), for every $\cP\cT$-symmetric scattering system,
    \[M_{11}(k)=\det\bM(k)M_{22}(k)^*.\]
This proves the following theorem.
    \begin{theorem}
    Spectral singularities of every $\cP\cT$-symmetric scattering system are self-dual.
    \label{am-thm-SS-selfdual}
    \end{theorem}
This does not however exclude the possibility of having non-$\cP\cT$-symmetric systems with self-dual spectral singularities. Simple examples of the latter are examined in \cite{am-jpa-2012, am-kalozoumis-2017,konotop-2017}.

\section{Summary and concluding remarks}

Scattering of waves can be studied using a general framework where the asymptotic solutions of the relevant wave equation are plane waves. This point of view is analogous to the general philosophy leading to the $\bS$-matrix formulation of scattering in the late 1930's. In one dimension, the transfer matrix proves to be a much more powerful tool than the $\bS$-matrix. We have therefore offered a detailed discussion of the transfer matrix and used it to introduce and explore the implications of $\cP$-, $\cT$-, and $\cP\cT$-symmetry. This is actually quite remarkable, for we could derive a number of interesting and useful quantitative results regarding the consequences of such symmetries without actually imposing them on the wave equation. These results apply to scattering phenomena modeled using local as well as nonlocal potentials and point interactions. The general setup we offer in Sec.~1 can also be used in the study of the scattering of a large class of nonlinear waves that are asymptotically linear. The results we derived using the transfer matrix may not however extend to such waves, because a useful nonlinear analog of the transfer matrix is not available. 

The recent surge of interest in the properties of $\cP\cT$-symmetric scattering potentials has led to the study of remarkable effects such as unidirectional invisibility, optical spectral singularities, and coherent perfect absorption. The global approach to scattering that we have outlined here allows for a precise description of these concepts for a general class of  scattering systems that cannot be described using a local scattering potential. In particular, we have derived specific conditions imposed by $\cP$-, $\cT$-, and $\cP\cT$-symmetry on the presence of nonreciprocal transmission and reflection, spectral singularities and their time-reversal, and unidirectional reflectionlessness and invisibility. 

A recent development that we have not covered in the present text is the construction of a transfer matrix for potential scattering in two and three dimensions \cite{am-pra-2016}. This has led to the discovery of a large class of exactly solvable multidimensional scattering potentials \cite{pra-2017}, and allowed for the extension of the  notions of spectral singularity and unidirectional invisibility to higher dimensions \cite{am-pra-2016,prsa-2016}. A particularly remarkable application of the multidimensional transfer matrix is the construction of scattering potentials in two dimensions that display perfect broadband invisibility below a tunable critical frequency \cite{ol-2017}.

\subsection*{Acknowledgments}

The author is indebted to Keremcan Do\u{g}an, Sasan HajiZadeh, and Neslihan Oflaz for their help in locating a large number of typos in the first draft of the manuscript.  This work has been supported by the Scientific and Technological Research Council of Turkey (T\"UB\.{I}TAK) in the framework of the project no: 114F357 and by Turkish Academy of Sciences (T\"UBA).

\ed
\begin{thebibliography}{99}

\bibitem{am-ahmed} Z.~Ahmed, New features of scattering from a one-dimensional non-Hermitian (complex) potential,  J.~Phys.\ A: Math.\ Theor.\ {\bf 45}, 032004 (2012).

\bibitem{bender-1998} C.~M.~Bender and S.~Boettcher, Real spectra in non-Hermitian Hamiltonians having $\cP\cT$ symmetry, Phys.\ Rev.\ Lett.~{\bf 80}, 5243-5246 (1998).

\bibitem{am-blashchak-1968} V.~A.~Blashchak, An analog of the inverse problem in
the scattering for a non-self-conjugate operator I, J.\ Diff.\ Eq.\ {\bf 4}, 1519-1533 (1968).

\bibitem{born-wolf} M.~Born and E.~Wolf, {\em Principles of Optics,} Cambridge University Press, Cambridge, 1999.

\bibitem{am-boyce-diPrima} W.~E.~Boyce and R.~C.~DiPrima, {\em Elementary Differential
Equations and Boundary Value Problems}, 10th Edition, Wiley, Hoboken, N.~J., 2012.

\bibitem{am-chong-2010} Y.~D.~Chong, L.~Ge, H.~Cao, and A.~D.~Stone, Coherent perfect absorbers: Time-reversed lasers, Phys.\ Rev.\ Lett.\ {\bf 105}, 053901 (2010).

\bibitem{am-chong-2011} Y.~D.~Chong, L.~Ge, and A.~D.~Stone, $\cP\cT$-symmetry breaking and laser-absorber modes in optical scattering systems, Phys.\ Rev.\ Lett.\ {\bf 106}, 093902 (2011).

\bibitem{devillard} P.~Devillard and B.~Souillard, Polynomially decaying transmission for the
nonlinear Schr\"odinger equation in a random medium, J.~Stat.\ Phys.\ {\bf 43}, 423-439 (1986).

\bibitem{p136} K.~Do\u{g}an, A.~Mostafazadeh, and M.~Sar{\i}saman, Spectral singularities, threshold gain, and output intensity for a slab laser with mirrors, preprint arXiv: 1710.02825, to appear in Ann. Phys. (N.Y.).

\bibitem{am-flugge} S.~Fl\"ugge, {\em Practical Quantum Mechanics}, Springer, Berlin, 1999.

%\bibitem{am-ge-2011} L.~Ge, Y.~D.~Chong, S.~Rotter, H.~E.~T\"ureci, and A.~D.~Stone, Phys.\ Rev.\ A~\textbf{84}, 023820 (2011).

\bibitem{stone-2012} L.~Ge, Y.~D.~Chong, and A,~D.~Stone, Conservation relations and anisotropic transmission resonances in one-dimensional $\cP\cT$-symmetric photonic heterostructures, Phys.\ Rev.~A {\bf 85} 023802 (2012).

\bibitem{am-jo-2017} H.~Ghaemi-Dizicheh, A.~Mostafazadeh, and M.~Sar{\i}saman, Nonlinear spectral singularities and laser output intensity, J.~Opt.~{\bf 19}, 105601 (2017).

\bibitem{am-greenberg} M.~Greenberg and M.~Orenstein, Irreversible coupling by use of dissipative optics, Opt.\ Lett.~{\bf 29}, 451-453 (2004).

\bibitem{am-ghuseynov} G.~Sh.~Guseinov, On the concept of spectral singularities, Pramana J.~Phys.~{\bf 73}, 587-603 (2009).

\bibitem{am-horsley} S.\ A.\ R. Horsley, M.\ Artoni and G.\ C.\ La Rocca,  Spatial Kramers-Kronig relations and the reflection of waves, {Nature Photonics} \textbf{9}, 436-439 (2015).

\bibitem{am-horsley-longhi} S.\ A.\ R. Horsley and S.\ Longhi, One-way invisibility in isotropic dielectric optical media, Amer.\ J.~Phys.~{\bf 85}, 439-446 (2017).

\bibitem{am-BSC} C.~W.~Hsu, B.~Zhen, A.~D.~Stone, J.~D.~Joannopoulos, and M.~Solja\v{c}i\'{c},
Bound states in the continuum, Nature Rev.\ Materials {\bf 1}, 16048 (2016).

\bibitem{am-jones-2012} H.~F.~Jones, Analytic results for a PT-symmetric optical structure,
J.~Phys.~A: Math.\ Theor.\ {\bf 45}, 135306 (2012).

\bibitem{am-kalozoumis-2017} P.~A.~Kalozoumis, C.~V.~Morfonios, G.~Kodaxis, F.~K.~Diakonos, and P.~Schmelcher, Emitter and absorber assembly for multiple self-dual operation and directional transparency, Appl.\ Phys.\ Lett.\ {\bf 110}, 121106 (2017).

\bibitem{am-kay-moses} I.\ Kay and H.~E.~Moses, Reflectionless transmission through dielectrics and scattering potentials, { J.~App.\ Phys.}~\textbf{27,} 1503-1508 (1956).

\bibitem{am-kemp-1958} R.~R.~D.~Kemp, A singular boundary value problem for a
non-self-adjoint differential opeartor, Canadian J.~Math. \textbf{10}, 447-462
(1958).

\bibitem{konotop-2017} V.~V.~Konotop and D.~A.~Zezyulin, Phase transition through the splitting of self-dual
spectral singularity in optical potentials, Opt.\ Lett.\ 42, 5206-5209 (2017).

\bibitem{am-kulishov}  M.~Kulishov, J.~M.~Laniel, N.~Belanger, J.~Azana, and D.~V.~Plant, Opt.\ Exp.~{\bf 13}, 3068-3078 (2005).

\bibitem{lamb} G.~L.~Lamb, {\em Elements of Soliton Theory}, Wiley, New York, 1980.

\bibitem{am-lin-2011} Z.~Lin, H.~Ramezani, T.~Eichelkraut, T.~Kottos, H.~Cao, D.~N.~Christodoulides, Unidirectional invisibility induced by PT-symmetric periodic structures, Phys.\ Rev.\ Lett.~{\bf 106}, 213901 (2011).

\bibitem{am-gupta-2014} X.~Liu, S.~Dutta Gupta, and G.~S.~Agarwal, Regularization of the spectral singularity in $\cP\cT$-symmetric systems by all-order nonlinearities: Nonreciprocity and optical isolation, Phys. Rev. A {\bf 89}, 013824 (2014).

\bibitem{am-longhi-2010-cpa} S.~Longhi, Backward lasing yields a perfect absorber, Physics {\bf 3}, 61 (2010).

\bibitem{am-longhi-2010-CPA-laser} S.~Longhi, $\cP\cT$-symmetric laser absorber, Phys. Rev. A {\bf 82}, 031801 (2010).

\bibitem{am-longhi-JPA-2011} S.~Longhi, Invisibility in PT -symmetric complex crystals,
J.~Phys.~A: Math.\ Theor.\ {\bf 44}, 485302 (2011).

\bibitem{am-longhi-2015}  S.\ Longhi,  Wave reflection in dielectric media obeying spatial
Kramers-Kronig relations, {EPL} \textbf{112}, 64001  (2015).

\bibitem{am-pra-2016} F.~Loran and A.~Mostafazadeh, Transfer matrix formulation
of scattering theory in two and three dimensions, Phys.\  Rev.~A {\bf 93}, 042707 (2016).

\bibitem{prsa-2016} F.~Loran and A.~Mostafazadeh, Unidirectional invisibility and nonreciprocal transmission in two and three dimensions, Proc.\ R.~Soc.~A {\bf 472}, 20160250 (2016).

\bibitem{pra-2017} F.~Loran and A.~Mostafazadeh, Class of exactly solvable scattering potentials in two dimensions, entangled-state pair generation, and a grazing-angle resonance effect, Phys.\ Rev.~A~{\bf 96}, 063837 (2017).

\bibitem{ol-2017} F.~Loran and A.~Mostafazadeh, Perfect broad-band invisibility in isotropic media with gain and loss, Opt.\ Lett.~{\bf 42}, 5250-5253 (2017).

\bibitem{am-messiah} A.~Messiah, {\em Quantum Mechanics}, Dover, New York, 1999.

\bibitem{am-jmp-2004} A.~Mostafazadeh, Pseudounitary operators and pseudounitary quantum dynamics, J.~Math.\ Phys.\ {\bf 45}, 932-946 (2004).

\bibitem{am-jpa-2006} A.\ Mostafazadeh, Delta-function potential with a complex coupling, J.~Phys.~A: Math.\ Gen.~{\bf 39}, 13495-13506 (2006).

\bibitem{am-prl-2009} A.~Mostafazadeh, Spectral singularities of complex scattering
 potentials and infinite reflection and transmission coefficients at real
 energies,  Phys.\ Rev.\ Lett.~\textbf{102}, 220402 (2009).

\bibitem{am-review} A.~Mostafazadeh, Pseudo-Hermitian representation of quantum mechanics,
Int.\ J.~Geom.\ Meth.\ Mod.\ Phys.\ {\bf 7}, 1191-1306 (2010).

\bibitem{am-pra-2011a} A.~Mostafazadeh, Optical spectral singularities as threshold resonances, Phys.\ Rev.~A {\bf 83}, 045801 (2011).

\bibitem{am-jpa-2011} A.~Mostafazadeh, Spectral singularities of a general point interaction,Ó J.~Phys.~A: Math.\ Theor.\ {\bf 44}, 375302 (2011).

\bibitem{am-jpa-2012} A.~Mostafazadeh, Self-dual spectral singularities and coherent perfect
absorbing lasers without $\cP\cT$-symmetry, J.~Phys.~A: Math.\ Gen.~{\bf 45}, 444024 (2012).

\bibitem{am-pra-2013} A.~Mostafazadeh, Invisibility and $\cP\cT$-symmetry, Phys.\ Rev.~A {\bf 87}, 012103 (2013).

\bibitem{am-prl-2013} A.~Mostafazadeh, Nonlinear spectral singularities for confined
nonlinearities, Phys.\ Rev.\ Lett.\ {\bf 110}, 260402 (2013).

\bibitem{am-ptrsa-2013} A.~Mostafazadeh, Pseudo-Hermitian quantum mechanics with unbounded metric operators, Phil.\ Trans.\ R.~Soc.\ A {\bf 371}, 20120050 (2013).

\bibitem{am-pra-2014a} A.~Mostafazadeh, Transfer matrices as non-unitary S-matrices, multimode
unidirectional invisibility, and perturbative inverse scattering, Phys.\ Rev.~A {\bf 89}, 012709 (2014).

\bibitem{am-pra-2014b} A.~Mostafazadeh, Unidirectionally invisible potentials as local building blocks of all scattering potentials, Phys.\ Rev.~A {\bf 90}, 023833 (2014).

\bibitem{am-jpa-2014c} A.~Mostafazadeh, Generalized unitarity and reciprocity relations for $\cP\cT$-symmetric scattering potentials, J.~Phys.~A: Math.\ Theor.\ {\bf 47},  505303 (2014).

\bibitem{am-SS-review} A.~Mostafazadeh, Physics of Spectral Singularities, in Proceedings of XXXIII Workshop on Geometric Methods in Physics, held in Bialowieza, Poland, June 29-July 5, 2014, Trends in Mathematics, pp.~145-165, Springer International Publishing, Switzerland, 2015; preprint arXiv:1412.0454.

\bibitem{am-ap-2016a} A.~Mostafazadeh, Point interactions, metamaterials, and $\cP\cT$-symmetry,  Ann.\ Phys.\ (NY) {\bf 368}, 56-69 (2016).

\bibitem{am-jpa-2016} A.~Mostafazadeh, Dynamical theory of scattering, exact unidirectional invisibility, and truncated $\fz~e^{2ik_0x}$ potential, J.~Phys.~A: Math.\ Theor.\ {\bf 49} 445302 (2016).

\bibitem{p145} A.~Mostafazadeh, Generalized unitarity relation for linear scattering systems in one dimension, preprint arXiv: 1711.04003.

\bibitem{pla-2017} A.~Mostafazadeh amd N.~Oflaz, Unidirectional reflection and invisibility in nonlinear media withanincoherent nonlinearity, Phys.\ Lett.~A {\bf 381}, 3548-3552 (2017).

\bibitem{am-pra-2012} A.~Mostafazadeh and S.~Rostamzadeh,  Perturbative analysis of spectral singularities and their optical realizations, Phys.\ Rev.~A {\bf 86}, 022103 (2012).

\bibitem{am-pla-2011} A.~Mostafazadeh and M.~Sar{\i}saman, Spectral singularities of a complex spherical barrier potential and their optical realization, Phys.\ Lett.~A {\bf 375}, 3387-3391 (2011).

\bibitem{am-prsa-2012} A.~Mostafazadeh and M.~Sar{\i}saman,
Optical spectral singularities and coherent perfect absorption in a two-layer spherical medium,
Proc.\ R.~Soc.~A {\bf 468}, 3224-3246 (2012).

\bibitem{am-pra-2013b} A.~Mostafazadeh and M.~Sar{\i}saman,
Spectral singularities and whispering gallery modes of a cylindrical gain medium,
Phys.\ Rev.~A {\bf 87}, 063834 (2013).

\bibitem{am-pra-2013d} A.~Mostafazadeh and M.~Sar{\i}saman,
Spectral singularities in the surface modes of a spherical gain medium,
Phys.\ Rev.~A {\bf 88}, 033810 (2013).

\bibitem{am-muga} J.~G.~Muga, J.~P.~Palao, B.~Navarro, and I.~L.~Egusquiza, Complex absorbing potentials, Phys.\ Rep.~{\bf 395}, 357-426 (2004).

\bibitem{am-Naimark} M.~A.~Naimark, Investigation of the spectrum and the expansion in eigenfunctions of a non-selfadjoint differential operator of the second order on a semi-axis, Am.\ Math.\ Soc.\ Transl.~{\bf 16} 103-193 (1960).\footnote{This is the English translation of M.~A.~Naimark, Trudy Moscov.\ Mat.\ Obsc.\ {\bf 3} 181-270 (1954).}

\bibitem{am-poladian} L.~Poladian, Resonance mode expansions and exact solutions for nonuniform gratings, Phys.\ Rev.\ E~{\bf 54}, 2963-2975 (1996).

\bibitem{prugovecki} E.~Prugove$\check{\rm c}$ki, {\em Quantum Mechanics in Hilbert Space,} Academic Press, New Yoek, 1981.

\bibitem{am-razavy} M.~Razavy, {\em Quantum Theory of Tunneling}, World Scientific, Singapore, 2003.

\bibitem{am-reed-simon} M.~Reed and B.~Simon, {\em Methods of Modern Mathematical Physics: Vol.~1 Functional Analysis}, Academic Press, San Diego, 1980.

\bibitem{ruschhaupt} A.~Ruschhaupt, T.~Dowdall1, M.~A.~Sim\'on, and J.~G.~Muga, ``Asymmetric scattering by non-Hermitian potentials,'' EPL {\bf 120}, 20001 (2017).

\bibitem{am-sanchez-soto} L.~ L.~S\'{a}nchez-Soto, J.~J.~Monz\'{o}na, A.~G.~ Barriuso, and J.~ F.~Cari\~{n}ena, The transfer matrix: A geometrical perspective, Phys.\ Rep.\ {\bf 513}, 191-227 (2012).

\bibitem{mustafa-2017} M.~Sar{\i}saman, Unidirectional reflectionlessness and invisibility in the TE and TM modes
of a PT -symmetric slab system, Phys.\ Rev.~A {\bf 95}, 013806 (2017).

\bibitem{am-schechter} M.~Schechter, {\em Operator Methods in Quantum Mechanics}, Dover, New York, 2002.

\bibitem{am-schwartz} J.~Schwartz,  Some non-selfadjoint operators, Commun.\ Pure.\  Appl.\  Math.~{\bf 13}, 609-639 (1960).

\bibitem{am-seigert-1939} A.~J.~F.~Seigert, On derivation of the dispersion formula for nuclear reactions, Phys.\ Rev.\ {\bf 56}, 750-752 (1939).

\bibitem{am-silfvast} W.~T.~Silfvast,  {\em Laser Fundamentals}, Cambridge University Press,  Cambridge, 1996.

\bibitem{am-stillinger-1975} F.~H.~Stillinger and D.~R.~Herrick, Bound states in the continuum, Phys.\ Rev.~A {\bf 11}, 446-454 (1975).

\bibitem{am-tureci-2006} H.~E.~T\"ureci,  A.~D.~Stone, and B.~Collier,
Self-consistent multimode lasing theory for complex or random lasing media,
Phys.~Rev.~A {\bf 74}, 043822 (2006).

\bibitem{am-vu} P.~L.~Vu, Explicit complex-valued solutions of the Korteweg--deVries equation on the half-line and on the whole-line, Acta Appl.\ Math.~{\bf 49}, 107Ð149, (1997).

\bibitem{am-wan} W.~Wan, Y.~Chong, L.~Ge, H.~Noh, A.~D.~Stone, and H.~Cao,  Time-reversed lasing and interferometric control of absorption, Science {\bf 331}, 889-892 (2011).

\bibitem{am-weinberg} S.~Weinberg, {\em Quantum Theory of Fields, Vol.~I}, Cambridge University Press, Cambridge, 1995.

\bibitem{am-wong-2016} Z.~J.~Wong, Y.-L.~Xu, J.~Kim, K.~O'Brien, Y.~Wang, L.~Feng, and X.~Zhang, Lasing and anti-lasing in a single cavity, Nature Photonoics {\bf 10}, 796-801 (2016).

\end{thebibliography}
